\documentclass[sigconf]{acmart}
\usepackage{enumitem}
\usepackage{multirow}
\usepackage{subfigure}
\usepackage{bbm}
\usepackage{abraces}
\usepackage[ruled,vlined]{algorithm2e}
\usepackage{balance}

\newtheorem{definition}{Definition}
\newtheorem{theorem}{Theorem}
\newtheorem{corollary}{Corollary}

\AtBeginDocument{%
  }

\setcopyright{acmcopyright}
\copyrightyear{2018}
\acmYear{2018}
\acmDOI{XXXXXXX.XXXXXXX}

\acmConference[Conference acronym 'XX]{Make sure to enter the correct
  conference title from your rights confirmation emai}{June 03--05,
  2018}{Woodstock, NY}
\acmPrice{15.00}
\acmISBN{978-1-4503-XXXX-X/18/06}

\begin{document}

\title{How Powerful is Graph Filtering for Recommendation}

\author{Shaowen Peng}
\email{peng.shaowen@naist.ac.jp}
\affiliation{%
  \institution{NARA Institute of Science and Technology}
  \city{Nara}
  \country{Japan}
}

\author{Xin Liu}
\email{xin.liu@aist.go.jp}
\affiliation{%
  \institution{National Institute of Advanced Industrial Science and Technology}
  \city{Tokyo}
  \country{Japan}
}

\author{Kazunari Sugiyama}
\email{sugiyama-k@g.osaka-seikei.ac.jp}
\affiliation{%
  \institution{Osaka Seikei University}
  \city{Osaka}
  \country{Japan}
}

\author{Tsunenori Mine}
\email{mine@m.ait.kyushu-u.ac.jp}
\affiliation{%
  \institution{Kyushu University}
  \city{Fukuoka}
  \country{Japan}
}

\begin{abstract}
It has been shown that the effectiveness of graph convolutional network (GCN) for recommendation is attributed to the spectral graph filtering. Most GCN-based methods consist of a graph filter or followed by a low-rank mapping optimized based on supervised training. However, we show two limitations suppressing the power of graph filtering:~(1) Lack of generality. Due to the varied noise distribution, graph filters fail to denoise sparse data where noise is scattered across all frequencies, while supervised training results in worse performance on dense data where noise is concentrated in middle frequencies that can be removed by graph filters without training. (2) Lack of expressive power. We theoretically show that linear GCN (LGCN) that is effective on collaborative filtering (CF) cannot generate arbitrary embeddings, implying the possibility that optimal data representation might be unreachable. \par

To tackle the first limitation, we show close relation between noise distribution and the sharpness of spectrum where a sharper spectral distribution is more desirable causing data noise to be separable from important features without training. Based on this observation, we propose a generalized graph normalization (${\rm G^2N}$) with hyperparameters adjusting the sharpness of spectral distribution in order to redistribute data noise to assure that it can be removed by graph filtering without training. As for the second limitation, we propose an individualized graph filter (IGF) adapting to the different confidence levels of the user preference that interactions can reflect, which is proved to be able to generate arbitrary embeddings. By simplifying LGCN, we further propose a simplified graph filtering for CF (SGFCF)$\footnote{\url{https://github.com/tanatosuu/sgfcf}}$ which only requires the top-$K$ singular values for recommendation. Finally, experimental results on four datasets with different density settings demonstrate the effectiveness and efficiency of our proposed methods.       

\end{abstract}




\maketitle

\section{Introduction}
Personalized recommendations have been widely applied to e-commerce, social media platforms, online video sites, etc., and has been indispensable to enrich people's daily life by offering the items user might be interested in based on the data such as user-item interactions, reviews, social relations, temporal information, etc. Among various recommendation scenarios, we focus on collaborative filtering (CF), a fundamental task for recommender systems. Conventional CF methods such as matrix factorization (MF) \cite{koren2009matrix} characterizes users and items as low dimensional vectors and predict the rating via the inner product between the corresponding embedding vectors. Subsequent works replace the linear design of MF with other advanced algorithms such as neural networks~\cite{covington2016deep,he2017neural}, attention mechanisms \cite{chen2017attentive,kang2018self}, transformer \cite{sun2019bert4rec,li2023graph}, diffusion models~\cite{wang2023diffusion}, etc. to model complex user-item relations.\par

While the aforementioned advanced recommendation algorithms show superior non-linear ability to model user-item relations, their performance are unstable due to the data sparsity issue in recommendation datasets. Graph Convolutional Networks (GCNs) recently have shown great potential in recommender systems due to the ability of capturing higher-order neighbor signals that can augment the training data to alleviate the sparsity issue. Early GCN-based methods adapt classic GCNs such as vanilla GCN \cite{kipf2017semi} and GraphSage \cite{hamilton2017inductive} to recommendation \cite{wang2019neural,zheng2018spectral,ying2018graph}. Subsequent works empower GCN by incorporating other advanced algorithms such as contrastive learning \cite{wu2020self}, learning in hyperbolic space \cite{sun2021hgcf,chen2023heterogeneous}, disentangled representation learning \cite{wang2020disentangled}, etc., slimming GCN model architectures to improve efficiency and scalability \cite{chen2020revisiting,he2020lightgcn,peng2022svd}, and study the effectiveness of GCN \cite{shen2021powerful,peng2022less}.\par

Recent studies have shown that the effectiveness of GCN for recommendation is mainly attributed to the spectral graph filtering which emphasizes important features (\textit{i.e.,} low frequencies) and filters out useless information. Most existing graph filtering designs can be classified to two categories: (1) Repeatedly propagating the node embeddings across the graph where the embeddings are optimized based on supervisory signals. This type of methods is actually equivalent to a low pass filter followed by a low-rank mapping~\cite{he2020lightgcn,peng2022less,guo2023manipulating} (see Equation (\ref{linear_gcn})). (2) A simple graph filter without model training~\cite{shen2021powerful,liu2023personalized}. This type of methods only relies on the graph filters to denoise. We can see these two kinds of methods are actually contradictory: the type (2) methods implies that data noise is only distributed in certain frequencies that can be simply removed by graph filters without training, which is contrary to the type (1), that we need model training to further remove data noise. However, we show that neither of them are perfect by pointing out two limitation suppressing the power of graph filtering. Firstly, both of the two designs lack generality. We find that noise distribution varies on datasets with different densities. Particularly, graph filters show poor performance on sparse datasets where data noise is scattered across all frequencies as they denoise by masking certain frequencies while fail to remove intrinsic noise in the features. On the other hand, despite the superior ability of supervised training learning from data polluted by noise,  it results in worse performance on dense datasets on which the noise is concentrated in middle frequencies that can be simply removed by graph filters. Moreover, most effective GCN-based methods are basically linear GCNs (LGCNs) without non-linearity. We theoretically show that they are incapable of generating arbitrary embeddings with multi-dimensions, implying the possibility that they cannot generate desirable user/item representations and demonstrating the lack of expressive power.\par

To tackle the first limitation, we further show the close relation between noise distribution and the sharpness of spectral distribution, that a sharper distribution is more desirable on which noise and important features are separable by graph filtering without supervised training. Based on this observation, we propose a generalized graph normalization (${\rm G^2N}$) to adjust the sharpness of spectrum via hyperparameters. As a result, data noise is redistributed through ${\rm G^2N}$, making the graph filtering generalizable on datasets with different densities. To tackle the second limitation, we propose an individualized graph filter (IGF) which is proved to generate arbitrary embeddings. Specifically, considering interactions do not equally reflect user preference (\textit{i.e.,} more (less) similar of the interacted items implies a higher (lower) consistency between the user's future and past behaviour), the proposed IGF emphasizes different frequencies based on the distinct confidence levels of user preference that interactions can reflect. Finally, by simplifying LGCN,  we propose a simplified graph filtering for CF (SGFCF) only requiring the top-$K$ singular values. Our main contributions can be summarized as follows: :
\begin{itemize}[leftmargin=10pt]
\item We point out two limitations suppressing the power of graph filtering for recommendation: (1) The lack of generality due to the the performance inconsistency of graph filters and supervised training on data with different densities and (2) The lack of expressive power of LGCN that is effective for CF.   

\item We propose a generalized graph normalization to tackle the first limitation, which redistributes data noise by adjusting the sharpness of spectrum, enabling the graph filtering to denoise without training on datasets with different densities.

\item We propose an individualized graph filtering to adapt to distinct confidence levels of user preference that interactions can reflect. It is proved to generate arbitrary data representations thus solves the second limitation. 

\item Extensive experimental results on four datasets with different density settings demonstrate the efficiency and effectiveness of our proposed method. 
\end{itemize}  

\section{Preliminaries}
We first introduce a commonly used GCN-learning paradigm for CF. Given an interaction matrix with implicit feedbacks containing $|\mathcal{U}|$ users and $|\mathcal{I}|$ items: $\mathbf{R}\in \{0, 1\}^{\left | \mathcal{U} \right | \times \left | \mathcal{I} \right |}$, we can define a bipartite graph $\mathcal{G}=(\mathcal{V}, \mathcal{E})$, where the node set contains all users and items: $\mathcal{V}=\mathcal{U}+\mathcal{I}$, the edge set contains the user-item pairs with interactions: $\mathcal{E}=\mathbf{R}^+$ where $\mathbf{R}^+=\{r_{ui}=1|u\in\mathcal{U}, i\in\mathcal{I}\}$. For simplicity, we let $|\mathcal{U}| + |\mathcal{I}|=n$. Then, we can define the corresponding adjacency matrix $\mathbf{A}$ of $\mathcal{G}$ and formulate the updating rules as:
\begin{equation}
\mathbf{H}^{(l+1)}=\sigma\left( \mathbf{\hat{A}} \mathbf{H}^{(l)} \mathbf{W}^{(l+1)} \right),
\end{equation}    
where $\sigma(\cdot)$ is an activation function, $\mathbf{\hat{A}}$ is a symmetric normalized adjacency matrix defined as follows:
\begin{equation}
\mathbf{\hat{A}}=\begin{bmatrix}
 \mathbf{0}& \mathbf{\hat{R}}\\ 
\mathbf{\hat{R}}^T &\mathbf{0} 
\end{bmatrix},
\label{adjacency_matrix}
\end{equation}
where $\mathbf{\hat{R}}=\mathbf{D}^{\mbox{-}\frac{1}{2}}_U\mathbf{R}\mathbf{D}^{\mbox{-}\frac{1}{2}}_I$ is a normalized interaction matrix, $\mathbf{D}_U$ and $\mathbf{D}_I$ are matrices with diagonal elements representing user and item degrees, respectively. The initial state is the stacked user/item embeddings $\mathbf{H}^{(0)}=\mathbf{E}\in\mathbb{R}^{n\times d}$, where each user $u$ and item $i$ are represented as learnable low-dimensional vectors. $\mathbf{W}^{(l+1)}\in \mathbb{R}^{d\times d} $ is a linear transformation. It has been shown that the following  linear GCN (LGCN) removing the activation function and linear transformation is effective for CF \cite{chen2020revisiting,he2020lightgcn}, with the final representations $\mathbf{O}$ generated as:
\begin{equation}
\mathbf{O}= g\left( \mathbf{\hat{A}} \right)\mathbf{E},
\label{linear_gcn}
\end{equation}
where $g(\mathbf{\hat{A}})$ is usually defined as a polynomial graph filter:
\begin{equation}
g\left(\mathbf{\hat{A}}\right)=\sum_{l=0}^L \theta_l \mathbf{\hat{A}}^l=\mathbf{V}diag\left( \sum_{l=0}^L \theta_l \lambda_k^l \right)\mathbf{V}^T,
\label{linear_gcn_decompose}
\end{equation}
where $\{\theta_0,\cdots,\theta_L\}$ is a set of parameters; $\lambda_k$ and $\mathbf{V}$ are an eigenvalue and the stacked eigenvectors, respectively; $diag(\cdot)$ is a diagonalization operator. The rating is predicted as:
\begin{equation}
\hat{r}_{ui}=\mathbf{o}_u^T\mathbf{o}_i
\label{rating_predict}
\end{equation}

\begin{definition}
(Graph Frequency). Given the eigenvalue and -vector pairs $(\lambda_k, \mathbf{v}_k)$ of $\mathbf{\hat{A}}$ where $\lambda_k\in[-1,1]$, the graph frequency is defined based on the variation of a signal on the graph $\left\|\mathbf{v}_k-\mathbf{\hat{A}}\mathbf{v}_k \right\|=1-\lambda_k\in[0,2]$. We call components with small variations low frequencies, components with high variations high frequencies.  
\end{definition}
According to Definition 1, low (high) frequencies emphasize the similarity (dissimilarity) between nodes and their neighborhood. We can adjust the weight of different frequencies via $g(\lambda_k)$ as $g(\mathbf{\hat{A}})=\sum_k g(\lambda_k)\mathbf{v}_k \mathbf{v}_k^T$. It has been shown that low frequencies are significantly contributive to recommendation \cite{shen2021powerful,peng2022less}, and most GCN-based methods can be classified to two categories: (1) a low pass filter followed by a linear mapping and optimization (\textit{i.e.,} LGCN) and (2) a simple graph filter without training \cite{shen2021powerful}. The learning process can be formulated as follows: 
\begin{equation}
\mathcal{G}\xrightarrow{\text{Spectrum}} \mathbf{V}\in \mathbb{R}^{n\times n}\xrightarrow{g(\lambda_k)} \mathbf{V}^{(K)} \in \mathbb{R}^{n\times K} \xrightarrow{\mathbf{E}} \mathbf{O}\in\mathbb{R}^{n \times d},
\end{equation} 
where $\mathbf{V}^{(K)}$ is the top-$K$ low frequency components. Here, we raise two questions on existing works: (1) The underlying assumption of the second type of methods is that data noise is only concentrated in certain frequencies that does not pollute important features (\textit{i.e.,} $\mathbf{V}^{(K)}$), which is contrary to the first type of methods that further training is required to denoise $\mathbf{V}^{(K)}$. Since the two types of methods seem to be contradictory, are they general on different datasets? (2) Despite the effectiveness and simplicity of LGCN, it also has a weakened expressive power compared with vanilla GCN, is it capable of generating desirable data representations? We will answer the two questions in Section 3.

\begin{table}[]
\centering
\caption{The accuracy (nDCG@10) of SGF and LGCN  with different density settings.}
\scalebox{0.9}{
\begin{tabular}{cccccc}
\toprule
\textbf{Datasets}                   & \textbf{Methods}  & \textbf{80\%}                    & \textbf{60\%}                    & \textbf{40\%}                    & \textbf{20\%}                    \\ \hline
\multirow{4}{*}{\textbf{CiteULike}} & SGF               & \textbf{0.2267} & \textbf{0.2222} & 0.1661                           & 0.0859                           \\
                                    & LGCN, $d=64$  & 0.1610                           & 0.2023                           & 0.2120                           & 0.1267                           \\
                                    & LGCN, $d=128$ & 0.1699                           & 0.2067                           & 0.2157                           & 0.1347                           \\
                                    & LGCN, $d=256$ & 0.1701                           & 0.2110                           & \textbf{0.2204} & \textbf{0.1390} \\ \midrule
\multirow{4}{*}{\textbf{Pinterest}} & SGF               & \textbf{0.0816} & \textbf{0.1008} & 0.1106                           & 0.0629                           \\
                                    & LGCN, $d=64$  & 0.0707                           & 0.0912                           & 0.1204                           & 0.1317                           \\
                                    & LGCN, $d=128$ & 0.0706                           & 0.0924                           & 0.1218                           & 0.1338                           \\
                                    & LGCN, $d=256$ & 0.0699                           & 0.0928                           & \textbf{0.1240} & \textbf{0.1341} \\\bottomrule
\end{tabular}}
\label{acc_density}
\end{table}

\section{Analysis on Graph Filtering}
In this section, we evaluate existing graph filtering designs in terms of generality and expressive power. In the light of the effectiveness of LGCN for CF, our analysis is mainly based on LGCN and a simple graph filter (SGF) $g(\mathbf{\hat{A}})$. The difference between the two models lies in the necessity of model training.
\subsection{Generality}
\subsubsection{Performance Inconsistency under Different Densities} 
We first evaluate graph filtering on datasets with different densities. We change the data density by adjusting the training ratio $x\%$ (the remaining is used as the test set), where $x=\{80, 60, 40, 20\}\%$. We choose an extensively used setting for CF with $\theta_0=,\cdots,=\theta_L$ \cite{he2020lightgcn} where $g(\mathbf{\hat{A}})$ is a low pass filter. We compare LGCN and SGF on two datasets and report the results in Table \ref{acc_density}. We observe that the performance of SGF and LGCN consistently decrease and increase as the training data is sparser, respectively. Particularly, SGF tends to be effective on the dense data (\textit{e.g.,} $x=80\%$ and 60\%) and significantly outperforms LGCN, indicating the uselessness of model training. On the other hand, the positive effect of training can be identified on the sparse data (\textit{e.g.,} $x=20\%$ and 40\%) as LGCN shows better performance. In addition, despite the improvement brought by increasing the embedding size, the performance tends to converge and still underperforms SGF on the dense data.

\begin{figure} \centering 
\subfigure[$x=80$ on CiteULike.] {  
\includegraphics[width=0.46\columnwidth]{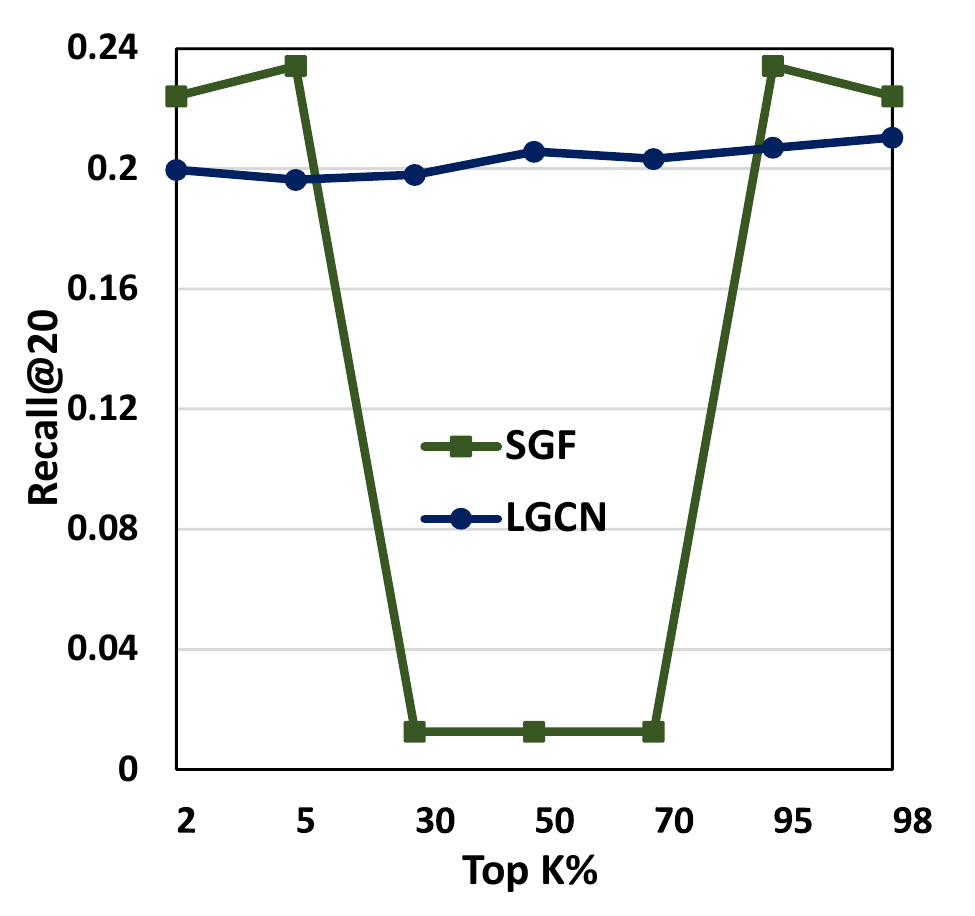} 
}
\subfigure[$x=20$ on CiteULike.] {  
\includegraphics[width=0.46\columnwidth]{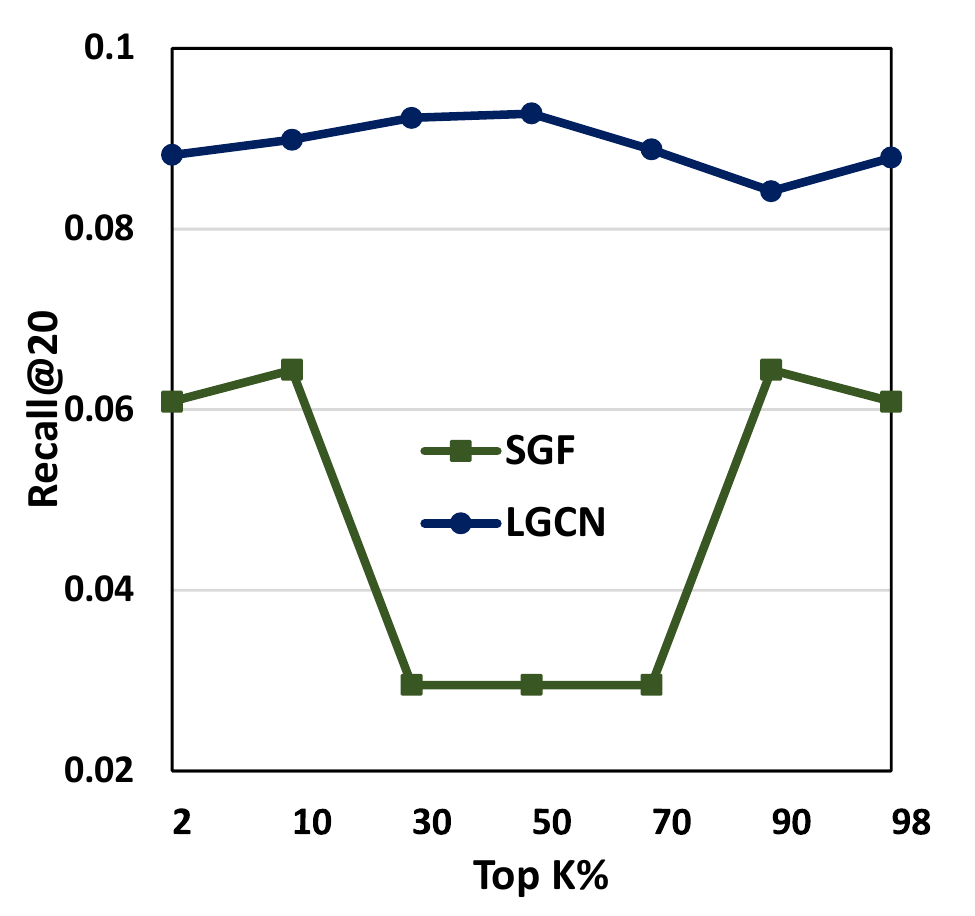} 
}
\subfigure[$x=80$ on Pinterest.] {  
\includegraphics[width=0.46\columnwidth]{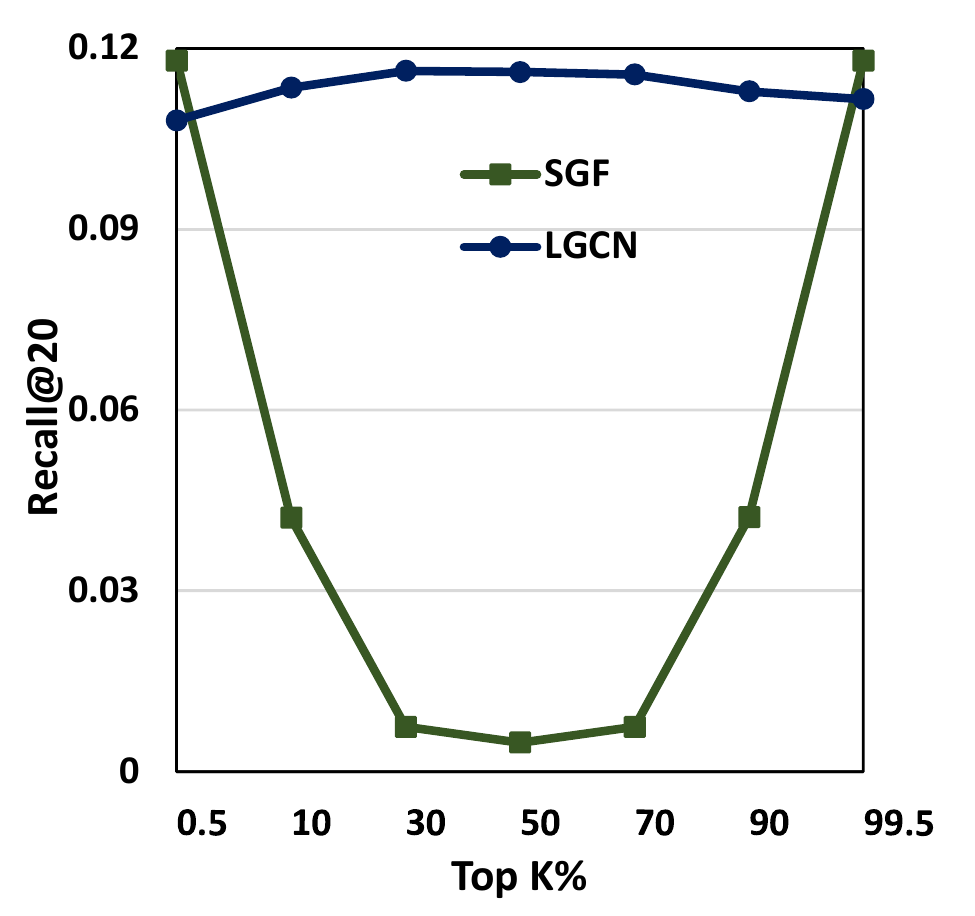} 
}
\subfigure[$x=20$ on Pinterest.] {  
\includegraphics[width=0.46\columnwidth]{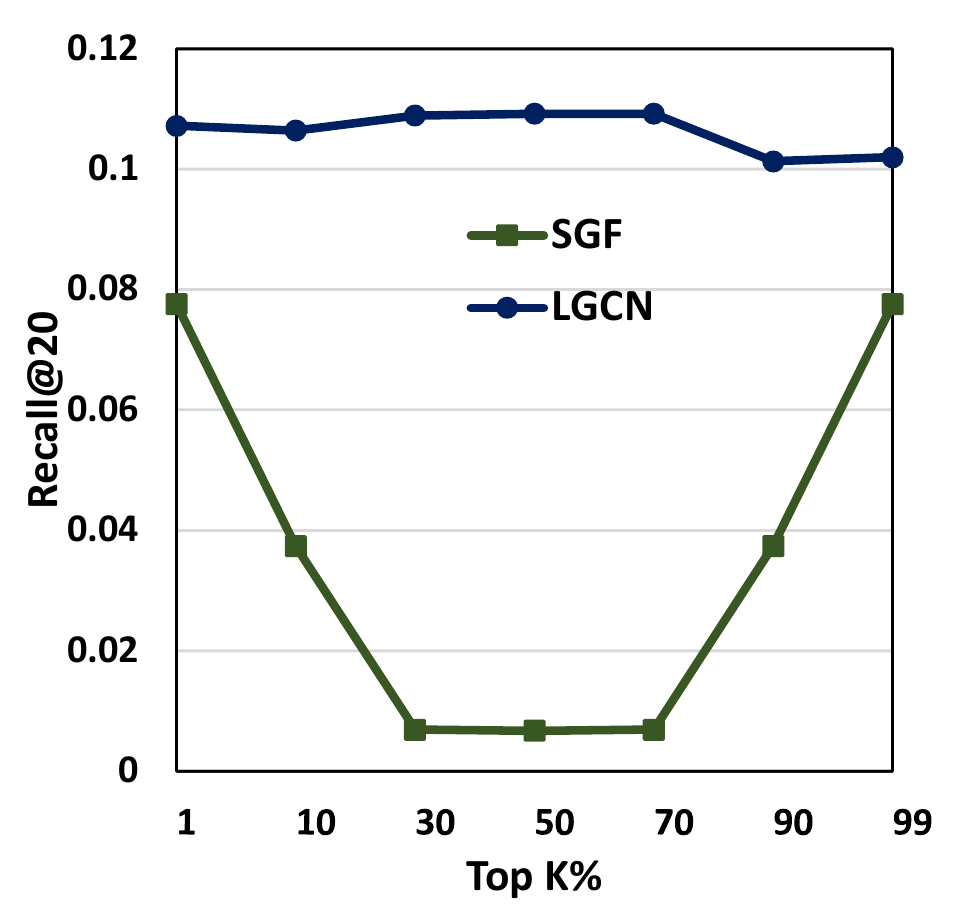} 
}
\caption{The accuracy (Recall@20) of SGF and LGCN when only considering top-$K$\% low frequencies.}           
\label{freq_noise}
\end{figure} 

\subsubsection{Noise Distribution Varies on Densities}
The above observations show the inconsistency and lack of generality of LGCN and SGF. Since SGF is a low pass filter, the poor performance on sparse data leads to a reasonable assumption that low frequencies might be polluted by noise. To verify this assumption, we conduct frequency analysis with the following filter:
\begin{equation}
g \left(\mathbf{\hat{A}} \right)=\mathbf{V}^{(K)}\mathbf{V}^{(K)^T},
\label{uniform_filter}
\end{equation}
We investigate the noisiness of certain frequencies by increasing $K$ from 0 and observing how accuracy changes after introducing certain frequencies. As the original polynomial filter $\sum_l^L \lambda_k^l$ emphasizes more on the lower frequencies leading to biased results, we choose a uniform filter here. We focus on $x=80\%$ and $x=20\%$ as SGF (LGCN) is most (least) and least (most) effective, respectively. We observe the followings from  the results shown in Figure \ref{freq_noise}:

\begin{itemize}[leftmargin=10pt]

\item On $x=80\%$ ((a) and (c)), the accuracy of SGF increases then significantly drops and rises again as $K$ increases, and the best performance outperforms LGCN, showing that the noise is concentrated in middle frequencies that can be removed by SGF. 

\item On $x=20\%$ ((b) and (d)), SGF underperforms LGCN as $K$ increases, indicating that the noise is distributed across all frequencies and pollutes the important features as well.

\item The stable performance of LGCN when incorporating noisy frequencies demonstrates the ability of model training of learning from noise. While it tends to be redundant and results in worse performance when the features are not polluted. 

\item The superior performance when only incorporating low frequencies shows that important graph features are distributed in low frequencies on both settings. 
\end{itemize}
The above observations verify our assumption that the poor performance of SGF on sparse data is attributed to the noise distribution that is across all frequencies. More importantly, both graph filters and supervised training lack generality due to their inconsistent performance on data with different densities, thus a more generic design is required. In addition, we notice that the performance of SGF is highly symmetric with respect to the middle frequency (\textit{i.e.,} $\lambda_{n/2}=0$) which is due to the following theorem and corollary:

\begin{theorem}
Given $\mathbf{P}$ and $\mathbf{Q}$ as the left and right singular vectors of $\mathbf{\hat{R}}$, we have the following relation:
\begin{equation}
\mathbf{V}=\begin{bmatrix}
 \mathbf{P}& \mathbf{P}\\ 
\mathbf{Q} &-\mathbf{Q} 
\end{bmatrix} / \sqrt{2}.
\end{equation}
Particularly, given a eigenvalue $\lambda_k>0$ with eigenvector $\mathbf{v}_k=[\mathbf{p}_k, \mathbf{q}_k]^T$, there always exists a $-\lambda_k$ with corresponding eigenvector $[\mathbf{p}_k, -\mathbf{q}_k]^T$. 
\end{theorem}

\begin{corollary}
Let $r_{ui}^{(K)}$ be the rating estimated based on the representation of Equation (\ref{uniform_filter}), we have the following relation:
\begin{equation}
r_{ui}^{(K)}= r_{ui}^{(n-K)}.
\end{equation}
\end{corollary}
Theorem and Corollary 1 show that low and high frequency are actually highly symmetric due to the bipartivity of $\mathcal{G}$. Therefore, we can only focus on the low and middle frequencies with $\lambda_k\geq0$.

\begin{figure} \centering 
\subfigure[] {  
\includegraphics[width=0.46\columnwidth]{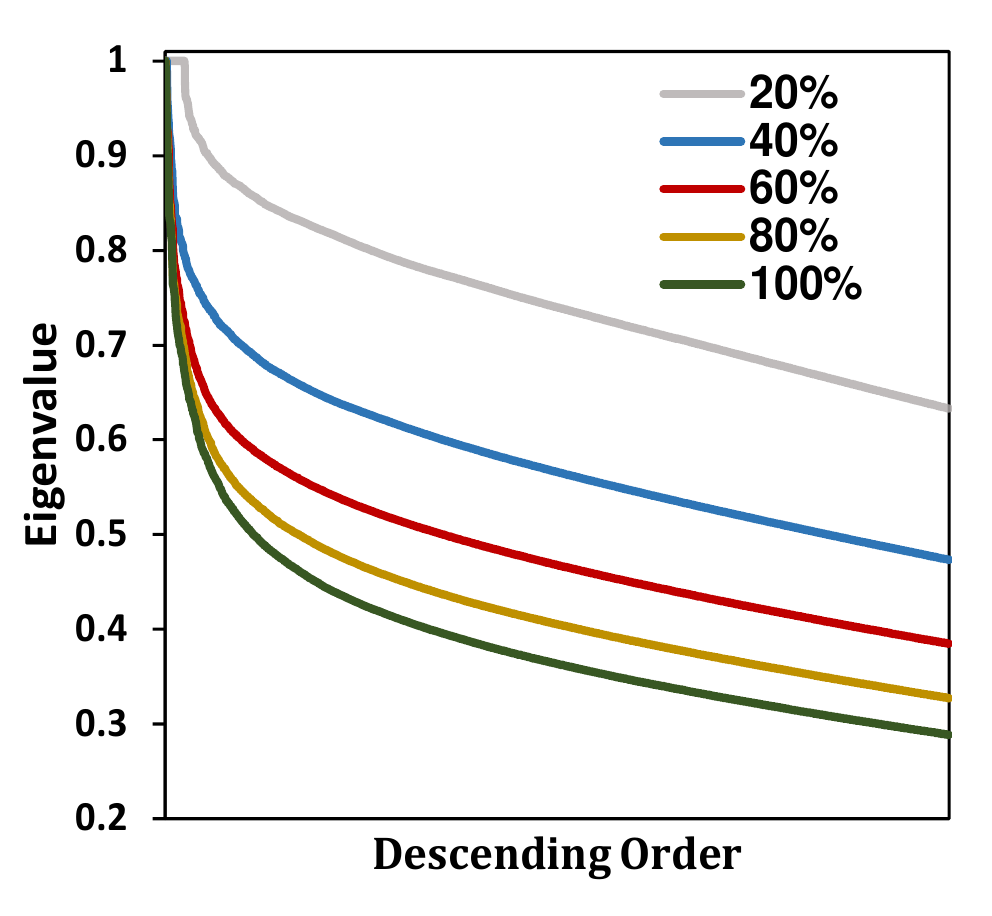} 
}
\subfigure[] {  
\includegraphics[width=0.46\columnwidth]{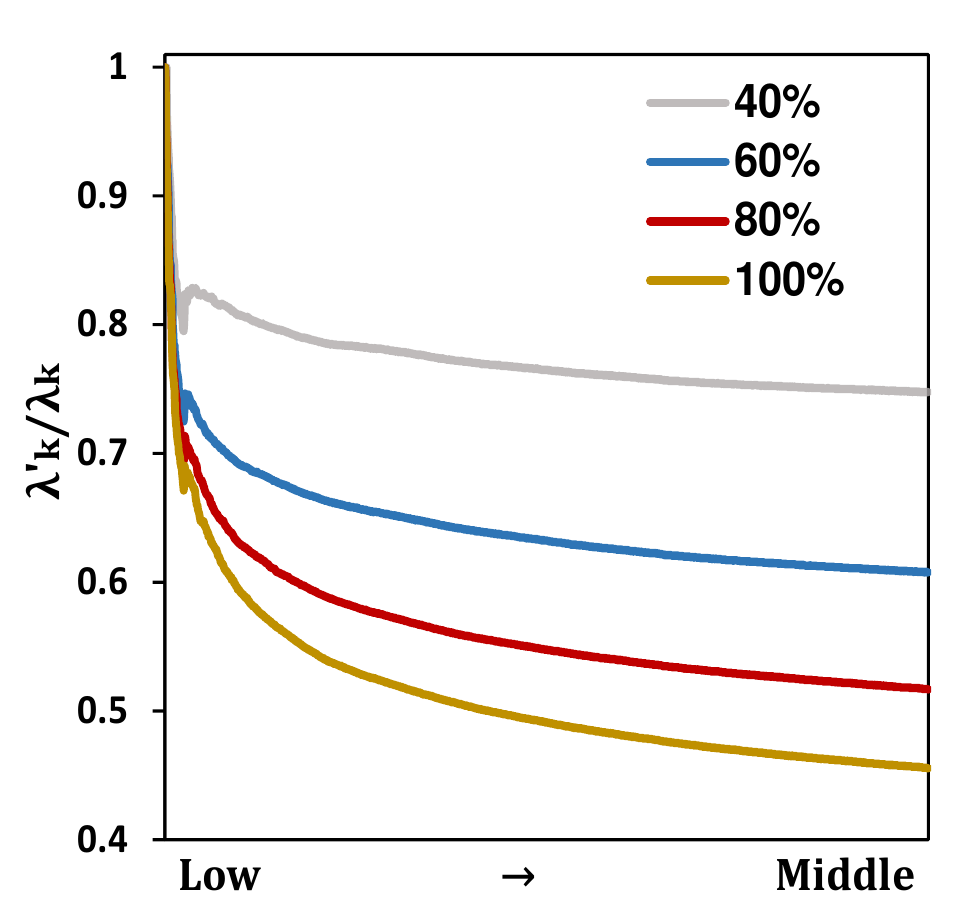} 
}

\caption{(a) Spectral distribution of CiteULike with different density settings. (b) the eigenvalue ratio ($\lambda'_k/\lambda_k$) of $x=\{40,60,80,100\}$ ($\lambda'_k$) to $x=20$ ($\lambda_k$). }           
\label{spectral_distribution}
\end{figure}

\subsection{Expressive Power}
Despite many existing works that can be summarized as LGCN variants\cite{he2020lightgcn,peng2022less,guo2023manipulating} choosing different filters to model user-item relations, its expressive power is still unknown.
\begin{theorem}
Assuming $\mathbf{\hat{A}}$ has no repeated eigenvalues, then LGCN can produce arbitrary embeddings when $d=1$. For $d>1$, each dimension requires an individual filter:
\begin{equation}
\mathbf{O}=\mathbf{V} \left(\mathbf{B}\Theta \right)\odot \mathbf{V}^T \mathbf{E},
\label{modified_lgcn}
\end{equation} 
where $\mathbf{B}\in \mathbb{R}^{n \times n}$ is a full-rank matrix with $\mathbf{B}_{kl}=\lambda_k^l$, $\Theta \in \mathbb{R}^{n \times d}$ is a parameter matrix. Equation (\ref{modified_lgcn}) degenerates to LGCN when $d=1$.
\end{theorem}
Theorem 2 shows that a shared global filter $[\theta_1,\cdots,\theta_n]$ cannot generate arbitrary multidimensional embeddings, and it is apparent one-dimensional representation is not enough to characterize users and items, which also demonstrates the incapability of LGCN to generate the optimal representations and shows the poor expressive power despite its effectiveness. Theorem 2 can be directly applied to SGF by setting $d=n$. Since the output has $n$ dimensions, it requires $n$ different filters to fit arbitrary embeddings.

\begin{figure} \centering 
\subfigure[] {  
\includegraphics[width=0.46\columnwidth]{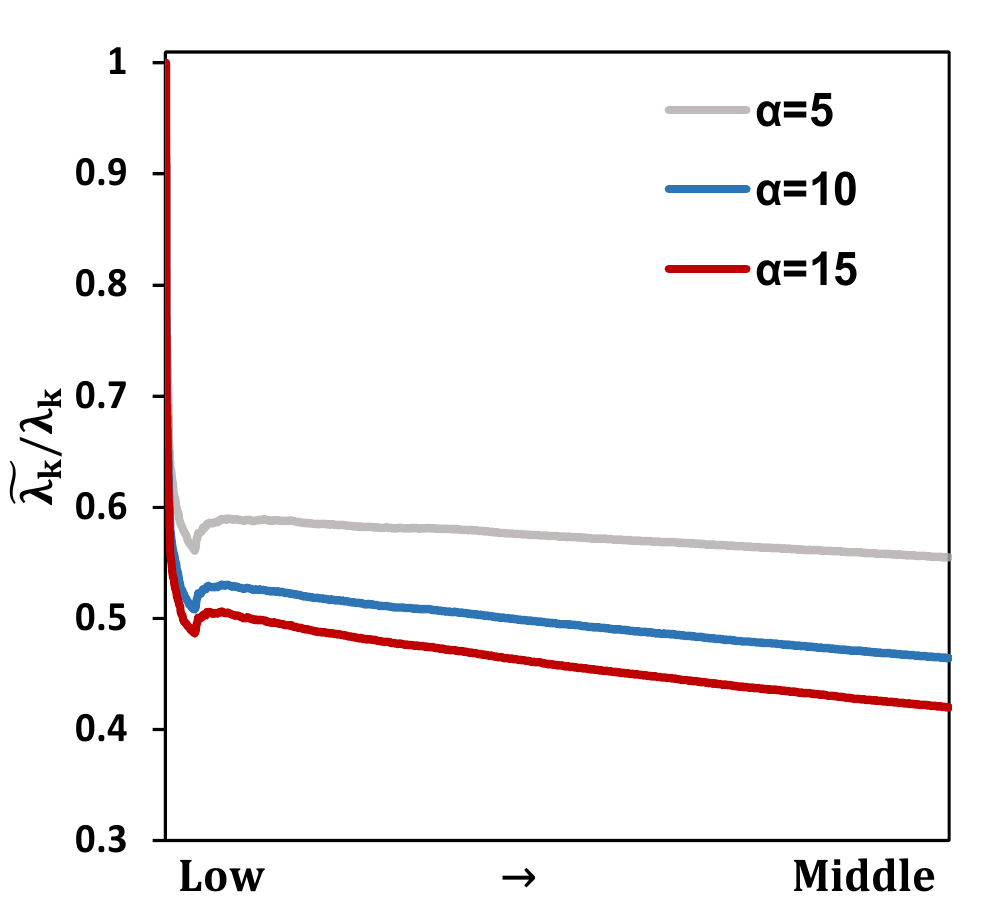} 
}
\subfigure[] {  
\includegraphics[width=0.46\columnwidth]{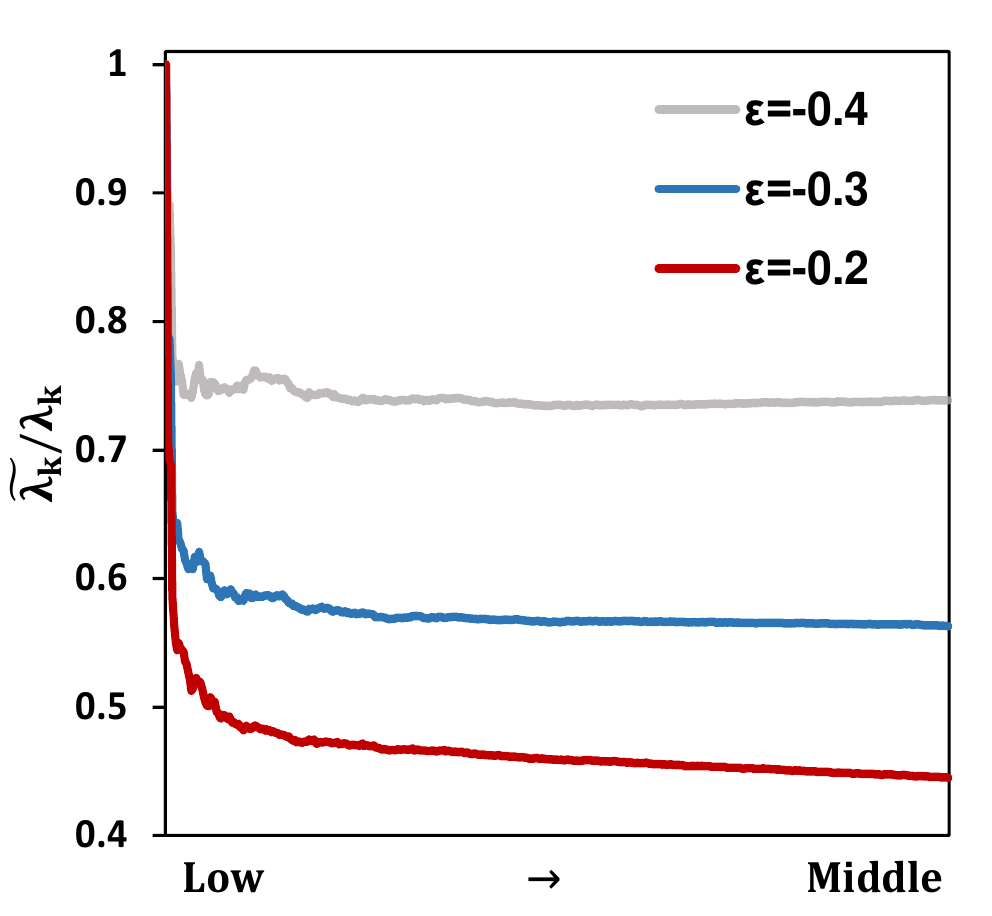} 
}

\caption{Eigenvalue ratio of $\mathbf{\tilde{A}}$ to $\mathbf{\hat{A}}$ with varying $\alpha$ and $\epsilon$ on CiteULike. }           
\label{graph_norm_effective}
\end{figure}

\section{Methodology}
\subsection{Sharpness of Spectrum Matters}
Ideally, if the spectrum is a low pass filter defined as follows:
\begin{equation}
\lambda_k=\left\{
             \begin{array}{lr}
             \gg 0 \quad  k\leq K    \\\\
             
             \approx 0 \quad \,  k>K ,\\ 
             \end{array}
\right.
\label{ideal_low}
\end{equation}
where $\lambda_1>\cdots>\lambda_{n/2}$. Most of the energy is concentrated in the top-$K$ low frequencies while the information distributed in the middle frequencies is trivial and noisy to data representations. In this case, the important features and noise can be separated by graph filtering without training. Inspired by this observation, we report the eigenvalue distribution with different density settings in Figure~\ref{spectral_distribution} (a). We observe that the denser data on which the noise and important features tend to be separable has a sharper spectrum. Particularly, in Figure \ref{spectral_distribution} (b), we can see that the eigenvalue closer to the middle frequency tends to drop faster than the low frequency. This means that the energy of middle frequencies is close to 0 while low frequencies stay important to data representation, causing the important features and noise to be distributed in different frequencies instead of mixed up together. Therefore, a sharper spectral distribution is more desirable as it is closer to an ideal low pass filter. Furthermore, consider a rank-$K$ approximation consisting of the top-$K$ low frequencies: $\mathbf{\hat{A}}_{K}=\sum_{k=1}^K \lambda_k \mathbf{v}_k \mathbf{v}_k^T$, then we have:
\begin{theorem}
Let $\mathbf{\tilde{A}}$ be a normalized adjacency matrix of $\mathcal{G}$ with a sharper spectral distribution than $\mathbf{\hat{A}}$: $1=\frac{\tilde{\lambda}_1}{\lambda_1}\geq  \cdots \geq \frac{\tilde{\lambda}_{K}}{\lambda_{K}}> \frac{\tilde{\lambda}_{K+1}}{\lambda_{K+1}}\geq\cdots\geq \frac{\tilde{\lambda}_{n/2}}{\lambda_{n/2}} $. Then $\mathbf{\tilde{A}}_{K}$ is a better approximation than $\mathbf{\hat{A}}_{K}$ with the following equation measuring the quality of an approximation:
\begin{equation}
\mathbf{appro} \left(\mathbf{\tilde{A}}_K \right)=\frac{\left|\mathbf{\tilde{A}}_{K} \right|_F^2} {\left|\mathbf{\tilde{A}} \right|_F^2},
\label{appro_measure}
\end{equation}
where $\left| \cdot \right|_F$ stands for the Frobenius norm.
\end{theorem}
Equation (\ref{appro_measure}) measures to what extent $\mathbf{\tilde{A}}_K $ can recover $\mathbf{\tilde{A}}$ where $\mathbf{appro} \left(\mathbf{\tilde{A}}_K \right) \in [0,1]$. As the spectral distribution becomes sharper, $\mathbf{\tilde{A}}_K $ is a better approximation indicating that more important information is concentrated in the top-$K$ low frequencies while the remaining information is more trivial that is distributed in the middle frequencies, causing them to be more separable.

\subsection{Generalized Graph Normalization ($\mathbf{G^2N}$)}
The analysis in Section 4.1 provides a solution to tackle the dilemma mentioned in Section 3.1, that graph filters and supervised training perform inconsistently on data with different density settings which is attributed to the varied noise distribution. If we are able to generate a desirable spectrum in a way that clearly differentiates important features from noise, it becomes possible to depend exclusively on graph filters for recommendation. Since the spectrum is closely related to how we normalize the graph, we study what graph normalization leads to a desirable spectrum in this subsection. \par
According to Definition 1, we know that putting more emphasis on low frequencies leads to a higher similarity between nodes and neighborhood. Thus, it is reasonable to assume that more energy is concentrated in the low frequencies if we increase the similarity by modifying the graph normalization, leading to a sharper spectrum. Since $g(\lambda_k)$ shares the same eigenspace with $\lambda_k$, we can simply define the similarity on $g(\mathbf{\hat{A}})=\mathbf{\hat{A}}$:  
\begin{equation}
\begin{aligned}
&\mathbf{\hat{A}}_{u*}\sum_{v\in\mathcal{N}_u^2}\mathbf{\hat{A}}_{v*}^T=\left(\mathbf{V}_{u*}\odot \boldsymbol{\lambda}\right)\mathbf{V}^T \left(\left(\sum_v\mathbf{V}_{v*}\odot \boldsymbol{\lambda}\right)\mathbf{V}^T\right)^T\\
&=\left(\mathbf{V}_{u*}\odot \boldsymbol{\lambda}\right) \left(\sum_v\mathbf{V}_{v*}\odot \boldsymbol{\lambda}\right)^T,
\end{aligned}
\label{sim_spec_relation}
\end{equation}
where $\boldsymbol{\lambda}$ is a vector containing all eigenvalues, $\mathbf{V}_{u*}$ refers to the $u$-th row of $\mathbf{V}$, $\odot$ is the operator for element-wise multiplication, $\mathcal{N}_u^2$ stands for the second-order neighborhood as only the second-order neighbors have similarity with $u$ (there is no similarity between a user and an item).\par

\begin{definition}
(Variation on the second-order Graph). The variation of the eigenvectors on the second-order graph is defined as:
\begin{equation}
\left\|\mathbf{v}_k-\mathbf{\hat{A}}^2\mathbf{v}_k\right\|=1-\lambda_k^2 \in [0,1].
\end{equation}  
\end{definition}
\textbf{Interpretation of Equation (\ref{sim_spec_relation}).} Definition 2 measures the difference between the signal samples of eigenvectors at each node ($\mathbf{V}_{uk}$) and at its second-order neighbors ($\sum_v\mathbf{V}_{vk}$). Intuitively, $\mathbf{v}_k$ with $|\lambda_k|\rightarrow1$ implies that the nodes are similar to their second-order neighborhood: $|\mathbf{V}_{uk}-\sum_v\mathbf{V}_{vk}|\rightarrow0$, while the middle frequency with $|\lambda_k|\rightarrow0$ emphasizes the difference between them. Consider $\boldsymbol{\lambda}$ as a band-pass filter, if the node similarity increases, the components with $|\lambda_k|\rightarrow1$ and $|\lambda_k|\rightarrow0$ should be correspondingly emphasized and suppressed to make the equation hold, respectively, leading to a sharper spectrum. In other words, the sharpness of the spectral distribution is closely related to the average node similarity defined on the normalized adjacency matrix.\par

Then, our question is transformed to: how do we increase the node similarity through a new graph normalization? The original setting is defined as $\mathbf{\hat{A}}_{ui}=\frac{1}{\sqrt{d_u}\sqrt{d_i}}$. Here, we define a renormalized adjacency matrix with $\mathbf{\tilde{A}}_{ui}=w(d_u)w(d_i)$, and the average similarity between users and items can be defined as follows:
\begin{equation}
\begin{aligned}
&{\rm SIM}_U=\frac{\sum_{u,v\in \mathcal{U}}\mathbf{\hat{A}}_{u*}\mathbf{\hat{A}}_{v*}^T}{|\mathcal{U}|^2}=\sum_{i\in \mathcal{I}} 2w(d_i)^2 \frac{\sum_{u,v\in \mathcal{N}_i} w(d_u)w(d_v)}{|\mathcal{U}|^2},\\
&{\rm SIM}_I=\frac{\sum_{i,j\in \mathcal{I}}\mathbf{\hat{A}}_{*i}^T\mathbf{\hat{A}}_{*j}}{|\mathcal{I}|^2}=\sum_{u\in \mathcal{U}} 2w(d_u)^2 \frac{\sum_{i,j\in \mathcal{N}_u} w(d_i)w(d_j)}{|\mathcal{I}|^2},\\
\end{aligned}
\label{ave_sim_user_item}
\end{equation} 
where $\mathcal{N}_u$/$\mathcal{N}_i$ is the set of first-hop neighborhood of $u/i$. We can see that the user/item with a higher degree has a larger impact on the average similarity (proportional to $d_u^2/d_i^2$), leading to the conclusion that the higher weights over the high-degree nodes results in higher node similarity. Based on the original design, we can propose two new designs with higher weights over high-degree nodes: (1) $w(d_u)=\frac{1}{\sqrt{d_u+\alpha}}$, and (2) $w(d_u)=d_u^{\epsilon}$, and propose a generalized graph normalization as follows:
\begin{equation}
\mathbf{\tilde{A}}=\left(\mathbf{D}+\alpha\mathbf{I}\right)^{\epsilon}\mathbf{A}\left(\mathbf{D}+\alpha\mathbf{I}\right)^{\epsilon},
\label{graph_norm}
\end{equation}
where $\alpha\geq0$, $\epsilon\in[-0.5,0]$. $\mathbf{\tilde{A}}=\mathbf{\hat{A}}$ when $\alpha=0$ and $\epsilon=-0.5$.
\begin{theorem}
Given $\tilde{\lambda}_k$ as the eigenvalue of $\mathbf{\tilde{A}}$, we have:
\begin{equation}
d_{max}\left( d_{max}+\alpha\right)^{2\epsilon} \lambda_k \geq \tilde{\lambda}_k \geq d_{min}\left( d_{min}+\alpha\right)^{2\epsilon} \lambda_k ,
\end{equation}
where $d_{min}$ and $d_{max}$ are the min and max node degree, respectively.
\end{theorem}
Particularly, increasing $\alpha$ and $\epsilon$ shrinks and scales the eigenvalue, respectively, making the spectrum not normalized any more. As we focus on the sharpness of the spectrum, we normalize $\tilde{\lambda}_k$ and visualize $\tilde{\lambda}_k/\lambda_k$ to observe how our proposed ${\rm G^2N}$ adjusts the spectrum shown in Figure \ref{graph_norm_effective}. We observe that the eigenvalue closer to the middle frequency drops more quickly, while the one closer to the low frequency tends to remain unchanged, and such a trend is more obvious as $\alpha$ or $\epsilon$ increases. The results in Figure \ref{graph_norm_effective} indicate that ${\rm G^2N}$ can generate a desirable spectrum which is more equivalent to an ideal low pass filter assuring that data noise and important features are linearly separable without further training.

\subsection{Individualized Graph Filtering (IGF)}
Owing to ${\rm G^2N}$, we can only rely on graph filters for recommendation. By applying the results in Theorem 2, we can empower SGF via the following enhanced model which is capable of generating arbitrary embeddings:
\begin{equation}
\begin{aligned}
&\mathbf{O}=\mathbf{V}\left(\mathbf{B}\Theta\right) \odot\mathbf{V}^T
=\left(\mathbf{V}\odot 
\begin{bmatrix}
g_1(\lambda_k)\\
\vdots\\
g_n(\lambda_k)
\end{bmatrix}\right)
\left(
\mathbf{V}\odot 
\begin{bmatrix}
g_1(\lambda_k)\\
\vdots\\
g_n(\lambda_k)
\end{bmatrix}
\right)^T,\\
&\mathbf{B}\Theta=\mathbf{B}
\begin{bmatrix}
\Theta_1,\cdots,\Theta_n
\end{bmatrix}
=
\begin{bmatrix}
g_1^2(\lambda_k),\cdots,g_n^2(\lambda_k)
\end{bmatrix}.
\end{aligned}
\end{equation}
Without loss of generality, we assume the filter $\mathbf{B}\Theta$ is non-negative. Here, each row of $\mathbf{V}$ can be considered as a user/item feature vector, and $[g_1(\lambda_k),\cdots,g_n(\lambda_k)]^T$ is the corresponding individualized filters related to users/items. There are different ways to implement the individualized filter, we present our solution as follows.
\begin{definition}
a (an) user's/item's homophilic ratio measuring the similarity of the the past interactions is defined as follows:
\begin{equation}
\begin{aligned}
&\mathbf{homo}(u)=\frac{\sum_{i,j \in \mathcal{N}_u} \mathbbm{1}^{\mathcal{D}_{\mathcal{G}/u}(i,j)<\delta}}{|\mathcal{N}_u|^2},\\
&\mathbf{homo}(i)=\frac{\sum_{u,v \in \mathcal{N}_i} \mathbbm{1}^{\mathcal{D}_{\mathcal{G}/i}(u,v)<\delta}}{|\mathcal{N}_i|^2},
\end{aligned}
\end{equation}
where $\mathcal{D}_\mathcal{G}(\cdot,\cdot)$ is the graph distance\footnote{The number of edges in the shortest path between two nodes.}, $\mathcal{D}_{\mathcal{G}/u}(i,j)$ measures the distance between $i$ and $j$ which does not pass through $u$, and $\mathbbm{1}$ is an indicator function producing 1 if two nodes are close enough (\textit{i.e.,} $<\delta$).
\end{definition}
Intuitively, if a user's interactions are similar (\textit{i.e.,} high homophilic ratio), then it is possible his/her future behaviour is consistent with the past interactions. While it is hard to rely on a user's past interactions if they are quite different (low homophilic ratio). Given that different frequencies emphasize the similarity between nodes and neighborhood with different degrees, it is reasonable to implement the individualized filter based on the homophilic ratio. 
After evaluating multiple graph filters, we choose a monomial filter: $g(\lambda_k)=\lambda_k^{\beta}$, map the homophilic ratio to $[\beta_1, \beta_2]$ via a linear function,
where the min and max of the homophilic ratio is mapped to $\beta_1$ and $\beta_2$, respectively. Then, the individualized filter is implemented as $g_u(\lambda_k)=\lambda_k^{\beta_u}$, where $\beta_u\in[\beta_1, \beta_2]$ is determined by $\mathbf{homo}(u)$.

\begin{table}
\centering
\caption{Statistics of datasets.}
\begin{tabular}{lcccc}
\toprule
Datasets&\#User&\#Item &\#Interactions &Density\%\\
\midrule
CiteULike&5,551&16,981&210,537&0.223\\
Pinterest&37,501&9,836&1,025,709&0.278\\
Yelp&25,677&25,815&731,672&0.109\\
Gowalla&29,858&40,981&1,027,370&0.084\\
\bottomrule
\label{datasets}
\end{tabular}
\end{table}

\subsection{Simplified Graph Filtering for CF (SGFCF)}
\begin{theorem}
The prediction matrix of SGF can be formulated as follows:
\begin{equation}
\mathbf{O}_U \mathbf{O}_I^T=\left(\mathbf{P} diag\left( \psi(\sigma_k) \right) \right) \left(\mathbf{Q} diag\left( \omega(\sigma_k) \right) \right)^T,
\label{svd_sgf}
\end{equation}
where $\sigma_k$ is the singular value of $\mathbf{\hat{R}}$, $\psi(\sigma_k)=\sum_{l=\{0, 2, \cdots \}}\sigma_k^l$, $\omega(\sigma_k)=\sum_{l=\{1, 3, \cdots \}}\sigma_k^l$.
\end{theorem}
Here, $\psi(\sigma_k)$ and $\omega(\sigma_k)$ are equivalent to low pass filters making no difference. By applying ${\rm G^2N}$ and IGF, we can design our SGFCF by modifying Equation (\ref{svd_sgf}) as follows:  
\begin{equation}
\mathbf{O}_U \mathbf{O}_I^T=\left(\mathbf{\tilde{P}}^{(K)}\odot G(\tilde{\sigma}_k) \right) \left(\mathbf{\tilde{Q}}^{(K)}\odot G(\tilde{\sigma}_k) \right)+\gamma \mathbf{\tilde{R}}\mathbf{\tilde{R}}^T\mathbf{\tilde{R}},
\label{sgfcf}
\end{equation}
where $G(\tilde{\sigma}_k)=[g_1(\tilde{\sigma}_k),\cdots,g_n(\tilde{\sigma}_k)]^T$; $\mathbf{\tilde{P}}^{(K)}$ and $\mathbf{\tilde{Q}}^{(K)}$ are the top-$K$ left and right singular vectors corresponding to low frequencies, $\tilde{\sigma}_k$ is the singular value of $\mathbf{\tilde{R}}=(\mathbf{D}_U+\alpha\mathbf{I})^{\epsilon}\mathbf{R}(\mathbf{D}_I+\alpha\mathbf{I})^{\epsilon}$, respectively. In practice, we notice that non-low frequencies are not completely noisy, thus we add a term $\mathbf{\bar{R}}\mathbf{\bar{R}}^T\mathbf{\bar{R}}$ containing all frequencies.

\subsection{Discussion}
Compared with existing GCN-based methods, our proposed SGFCF mainly differs from them in three aspects: (1) We provide a closed-form solution with complexity only as: $\mathcal{O}(K\left|\mathbf{R}^+\right|+K^2\left|\mathcal{U}\right|+K^2\left|\mathcal{I} \right|)$. (2) Our method is generalizable on datasets with different densities. (3) Compared with the methods that can be summarized as LGCN, our method is proven to have stronger expressive power which is capable of generating arbitrary embeddings. Particularly, compared with non-parametric methods such as GFCF \cite{shen2021powerful}, our superiority comes from two aspects: (1) GFCF is equivalent to a low pass filter which does not consider the varied noise distribution on data with different densities, thus is not generalizable. (2) It can be summarized as a LGCN showing poor expressive power. We will empirically demonstrate our superiority in Section 5.

\section{Experiment} 

\begin{table*}[]
\caption{Performance comparison on CiteULike and Yelp. Improv.\% denotes the improvement over the best baseline.}
\scalebox{0.859}{
\begin{tabular}{cc|cccccccc}
\toprule
\multicolumn{2}{c|}{\multirow{2}{*}{}}               & \multicolumn{4}{c}{\textbf{CiteULike}}                                    & \multicolumn{4}{c}{\textbf{Yelp}}                                         \\
\multicolumn{2}{c|}{}                                & \multicolumn{2}{c}{\textbf{x=80\%}} & \multicolumn{2}{c}{\textbf{x=20\%}} & \multicolumn{2}{c}{\textbf{x=80\%}} & \multicolumn{2}{c}{\textbf{x=20\%}} \\
\multicolumn{2}{c|}{\textbf{Method}}                 & \textbf{nDCG}    & \textbf{Recall}  & \textbf{nDCG}    & \textbf{Recall}  & \textbf{nDCG}    & \textbf{Recall}  & \textbf{nDCG}    & \textbf{Recall}  \\ \midrule
\multirow{9}{*}{\textbf{Training-Based}} & BPR       & 0.1620           & 0.1778           & 0.0674           & 0.0621           & 0.0487           & 0.0607           & 0.0635           & 0.0600           \\
                                         & DirectAU  & 0.2102           & 0.2260           & 0.1348           & 0.1280           & \underline{ 0.0721}     & \underline{ 0.0872}     & 0.0857           & 0.0839           \\
                                         & LightGCN  & 0.1610           & 0.1777           & 0.1273           & 0.1206           & 0.0572           & 0.0721           & 0.0751           & 0.0725           \\
                                         & SGL-ED    & 0.1890           & 0.2117           & 0.1217           & 0.1167           & 0.0676           & 0.0837           & 0.0817           & 0.0784           \\
                                         & XSimGCL   & 0.2024           & 0.2229           & 0.1360           & 0.1289           & 0.0691           & 0.0847           & 0.0837           & 0.0809           \\
                                         & LightGCL  & 0.2096           & 0.2214           & 0.1270           & 0.1224           & 0.0673           & 0.0836           & 0.0682           & 0.0661           \\
                                         & DCCF      & 0.1641           & 0.1819           & 0.0791           & 0.0740           & 0.0626           & 0.0746           & 0.0668           & 0.0627           \\
                                         & GDE       & 0.1890           & 0.2055           & \underline{ 0.1518}     & \underline{ 0.1429}     & 0.0653           & 0.0805           & 0.0866           & 0.0839           \\
                                         & JGCF      & 0.1557           & 0.1752           & 0.1438           & 0.1386           & 0.0626           & 0.0789           & \underline{ 0.0895}     & \underline{ 0.0868}     \\ \midrule
\multirow{4}{*}{\textbf{Training-Free}}  & EASE      & 0.2368           & 0.2468           & 0.1313           & 0.1211           & 0.0719           & 0.0859           & 0.0360           & 0.0346            \\
                                         & GFCF      & \underline{ 0.2405}     & \underline{ 0.2562}     & 0.0902           & 0.0896           & 0.0644           & 0.0806           & 0.0722           & 0.0725           \\
                                         & BPSM      & 0.2333           & 0.2466           & 0.0886           & 0.0872           & 0.0622           & 0.0748           & 0.0113           & 0.0110           \\
                                         & PGSP      & 0.2357           & 0.2501           & 0.1121           & 0.1105           & 0.0643           & 0.0789           & 0.0732           & 0.0717           \\ \midrule
\multirow{2}{*}{\textbf{Ours}}           & SGFCF     & \textbf{0.2663}  & \textbf{0.2797}  & \textbf{0.1542}  & \textbf{0.1478}  & \textbf{0.0824}  & \textbf{0.0998}  & \textbf{0.0963}  & \textbf{0.0930}  \\
                                         & Improv.\% & +8.69            & +9.17            & +1.58            & +3.43            & +14.29           & +14.45           & +7.60            & +3.56            \\ \bottomrule
\end{tabular}}
\label{accuracy_compare1}
\end{table*}

\begin{table*}[]
\caption{Performance comparison on Pinterest and Gowalla. Improv.\% denotes the improvement over the best baseline.}
\scalebox{0.859}{
\begin{tabular}{cc|cccccccc}
\toprule
\multicolumn{2}{c|}{\multirow{2}{*}{}}               & \multicolumn{4}{c}{\textbf{Pinterest}}                                    & \multicolumn{4}{c}{\textbf{Gowalla}}                                      \\
\multicolumn{2}{c|}{}                                & \multicolumn{2}{c}{\textbf{x=80\%}} & \multicolumn{2}{c}{\textbf{x=20\%}} & \multicolumn{2}{c}{\textbf{x=80\%}} & \multicolumn{2}{c}{\textbf{x=20\%}} \\
\multicolumn{2}{c|}{\textbf{Method}}                 & \textbf{nDCG}    & \textbf{Recall}  & \textbf{nDCG}    & \textbf{Recall}  & \textbf{nDCG}    & \textbf{Recall}  & \textbf{nDCG}    & \textbf{Recall}  \\ \midrule
\multirow{9}{*}{\textbf{Training-Based}} & BPR       & 0.0668           & 0.0780           & 0.0976           & 0.0946           & 0.1164           & 0.1186           & 0.1086           & 0.0917           \\
                                         & DirectAU  & 0.0840           & 0.0964           & 0.1338           & 0.1289           & 0.1286           & 0.1349           & 0.1864           & 0.1648           \\
                                         & LightGCN  & 0.0699           & 0.0828           & 0.1297           & 0.1263           & 0.0987           & 0.1074           & 0.1477           & 0.1368           \\
                                         & SGL-ED    & 0.0755           & 0.0888           & 0.1355           & 0.1300           & 0.1343           & 0.1417           & 0.1789           & 0.1563           \\
                                         & XSimGCL   & 0.0842           & 0.0973           & 0.1396           & 0.1346           & 0.1288           & 0.1392           & 0.1861           & 0.1643           \\
                                         & LightGCL  & 0.0717           & 0.0826           & 0.1342           & 0.1287           & 0.1368           & 0.1436           & 0.1524           & 0.1329           \\
                                         & DCCF      & 0.0767           & 0.1002           & 0.1241           & 0.1151           & 0.1179           & 0.1181           & 0.1452           & 0.1244           \\
                                         & GDE       & 0.0748           & 0.0862           & \underline{ 0.1403}     & \underline{ 0.1351}     & 0.1261           & 0.1313           & \underline{ 0.1847}     & \underline{ 0.1645}     \\
                                         & JGCF      & 0.0765           & 0.0885           & 0.1363           & 0.1305           & 0.1173           & 0.1260           & 0.1845           & 0.1631           \\ \midrule
\multirow{4}{*}{\textbf{Training-Free}}  & EASE      & 0.0853           & 0.0964           & 0.1064           & 0.1020           & 0.1350           & 0.1440           & 0.1405           & 0.1247           \\
                                         & GFCF      & 0.0859           & 0.0981           & 0.0851           & 0.0857           & 0.1387           & \underline{ 0.1483}     & 0.1662           & 0.1474           \\
                                         & BPSM      & 0.0858           & 0.0950           & 0.0567           & 0.0558           & \underline{\textbf{ 0.1432}}     & 0.1482           & 0.0737           & 0.0694           \\
                                         & PGSP      & \underline{ 0.0876}     & \underline{ 0.1000}     & 0.1264           & 0.1219           & 0.1395           & 0.1462           & 0.1640           & 0.1450           \\ \midrule
\multirow{2}{*}{\textbf{Ours}}           & SGFCF     & \textbf{0.0923}  & \textbf{0.1052}  & \textbf{0.1437}  & \textbf{0.1384}  & \textbf{0.1432}  & \textbf{0.1527}  & \textbf{0.1973}  & \textbf{0.1762}  \\
                                         & Improv.\% & +5.36            & +4.80            & +2.42            & +2.44            & +0.00            & +2.97            & +5.85            & +6.92            \\ \bottomrule
\end{tabular}}
\label{accuracy_compare2}
\end{table*}

\subsection{Experimental Setup} 
\subsubsection{Datasets and Evaluation Metrics}
We evaluate our method on four datasets in this work, the statistics are summarized in Table \ref{datasets}. CiteULike\footnote{https://github.com/js05212/citeulike-a} is collected from a social bookmarking service CiteULike which allows users to bookmark and share research articles; Pinterest \cite{he2017neural} is constructed for evaluating content-based image recommendation; Yelp \cite{he2016fast} is from the Yelp Challenge data; Gowalla \cite{wang2019neural} is a check-in dataset which records the locations users have visited. We focus on $x=80\%$ and $x=20\%$, randomly select 5\% as validation set, and leave the remaining for test. We adopt Recall and nDCG \cite{jarvelin2002cumulated}, two widely used evaluation metrics for personalized recommendation. The recommendation list is generated by ranking unobserved items and truncating at position $k=10$.

\subsubsection{Implementation}
We use stochastic gradient descent (SGD) as the optimizer for training-based models. The embedding size $d$ is set to 64, the regularization rate $\gamma$ is set to 0.01 on all datasets, the learning rate is tuned with step size 0.1, the model parameters are initialized with Xavier initialization \cite{glorot2010understanding} and the batch size is set to 256. $\alpha\geq0$ and $\epsilon\in[-0.5,0]$ are tuned with step size 1 and 0.02, respectively. Other hyperparameters in this work are all tuned with step size 0.1. We set $\delta=2$ which is the smallest graph distance between homogeneous nodes; we use a monomial filter: $g(\lambda_k)=\lambda_k^{\beta}$, $\beta_1\leq \beta$ and $\beta_2\geq \beta$ are tuned after determining the best $\beta$.

\subsubsection{Baselines}
We compare our proposed methods with competitive baselines which can be categorized to training-based and non-parametric (\textit{i.e.,} training-free) methods. For training-based methods, we choose BPR \cite{rendle2009bpr} and DirectAU \cite{wang2022towards} implemented on MF, and seven GCN-based methods: LightGCN \cite{he2020lightgcn}, SGL-ED~\cite{wu2020self}, LightGCL~\cite{caisimple}, XSimGCL \cite{yu2023xsimgcl}, DCCF \cite{ren2023disentangled}, GDE \cite{peng2022less}, and JGCF~\cite{guo2023manipulating}. Additionally, we adopt four non-parametric methods: EASE \cite{steck2019embarrassingly}, GF-CF~\cite{shen2021powerful}, PGSP \cite{liu2023personalized}, and BPSM \cite{choi2023blurring}. Note that the hyperparameters are properly set after conducting tuning for them on the datasets used in this work.

\subsection{Comparison}
\subsubsection{Accuracy}
We report the accuracy of baselines and our method in Table \ref{accuracy_compare1} and \ref{accuracy_compare2}. We have the following observations:

\begin{itemize}[leftmargin=10pt]

\item Overall, GCN-based methods show better performance especially on sparse data, indicating the superior ability to tackle data sparsity by incorporating higher-order neighborhood. For instance, LightGCN underperforms BPR on two datasets with $x=80\%$ while significantly outperforms BPR on four datasets with sparse setting $x=20\%$.

\item Comparing non-parametric methods (\textit{i.e.,} Ease, GFCF, PGSP and BPSM) with training-based methods,  we can see that non-parametric methods tend to be effective on dense data (\textit{e.g.,} $x=80\%$) and show relatively poor performance on sparse data (\textit{e.g.,} $x=20\%$). For instance, GFCF achieves the best baseline on $x=80\%$ while shows the second worst performance on $x=20\%$ on CiteULike. The relatively poor performance of GFCF on sparse data also verifies our previous analysis in Section 3.1 that a simple low pass filter cannot work well on datasets with different densities due to the varied noise distribution.

\item Contrary to non-parametric methods, GCN-based methods requiring training such as LightGCN, GDE, and JGCF show superior performance on sparse data while are less effective on dense data, which further verifies the analysis in Section 3.1 showing the performance inconsistency of graph filters and supervised training on data with different densities.

\item Our proposed SGFCF, significantly outperforms competitive baselines almost across all datasets, demonstrating the effectiveness of our proposed method. Particularly, SGFCF outperforms GFCF which shares similarities with our designs by 14.0\% on $x=80\%$ and by 41.0\% on $x=20\%$ on average, in terms of nDCG@10. The larger improvement on sparser data further proves the superiority of our proposed designs over GFCF.

\end{itemize}

\begin{figure} \centering 
\subfigure[$x=80\%$ on Yelp.] {  
\includegraphics[width=0.46\columnwidth]{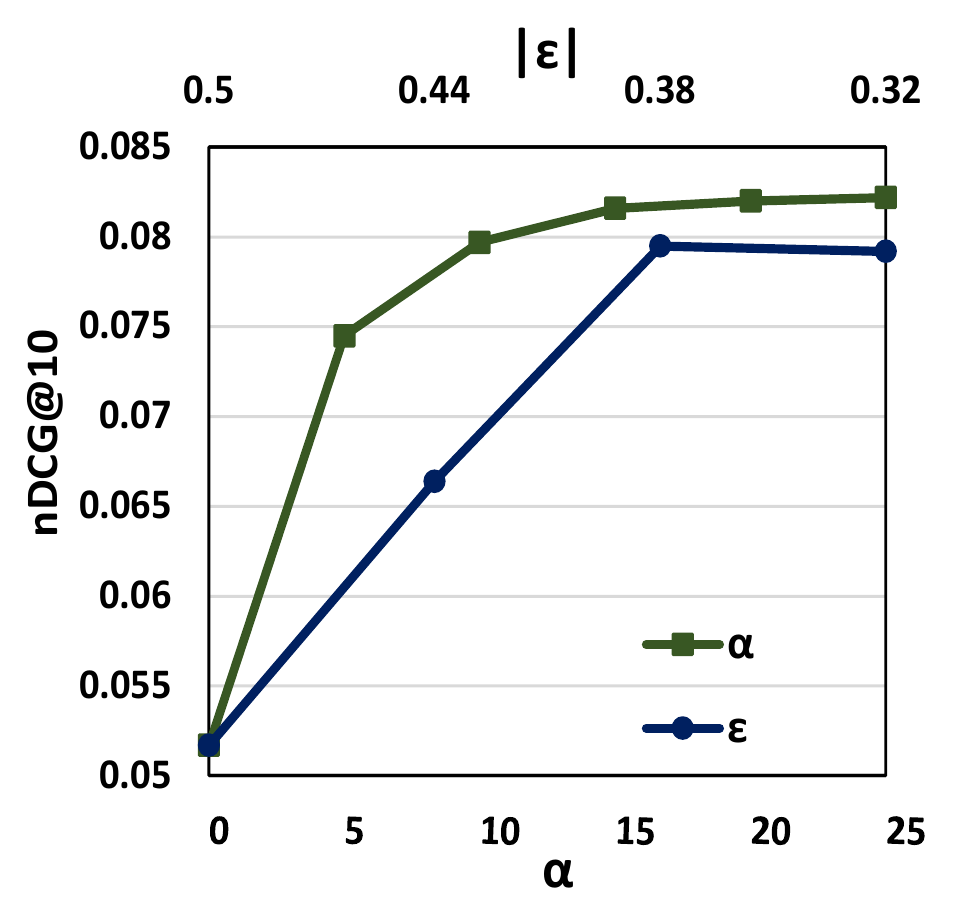} 
}
\subfigure[$x=20\%$ on Yelp.] {  
\includegraphics[width=0.46\columnwidth]{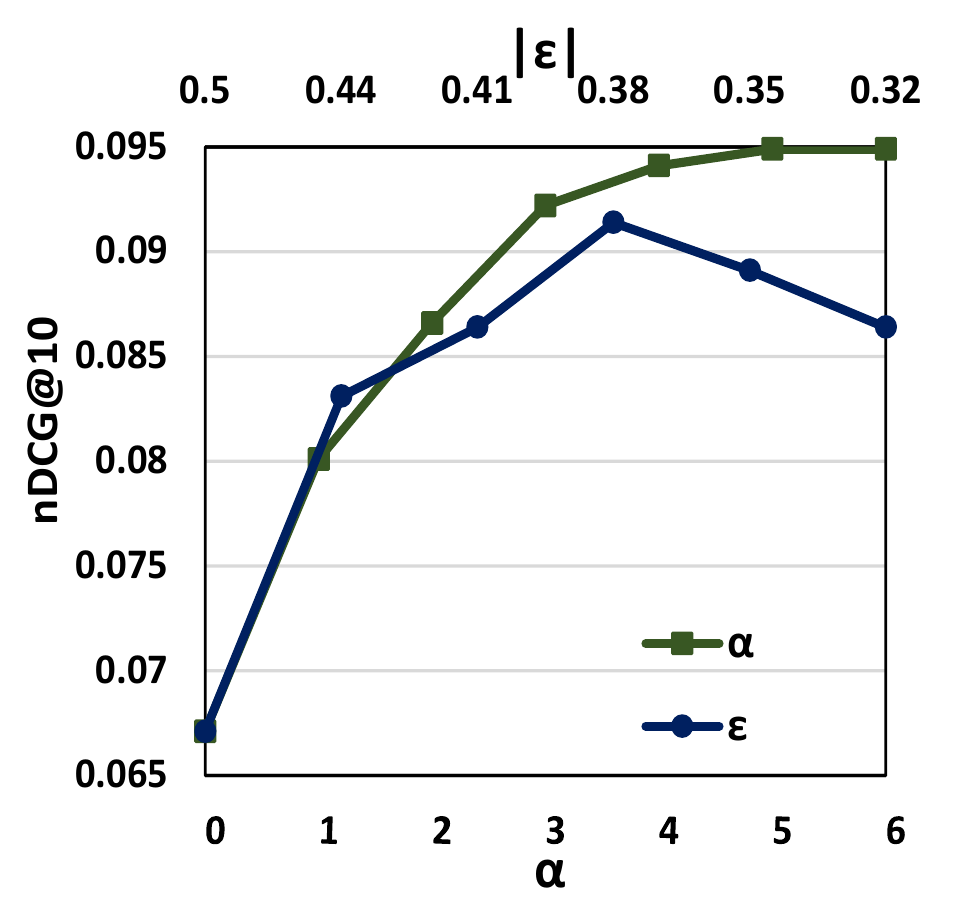} 
}
\subfigure[$x=80\%$ on CiteULike.] {  
\includegraphics[width=0.46\columnwidth]{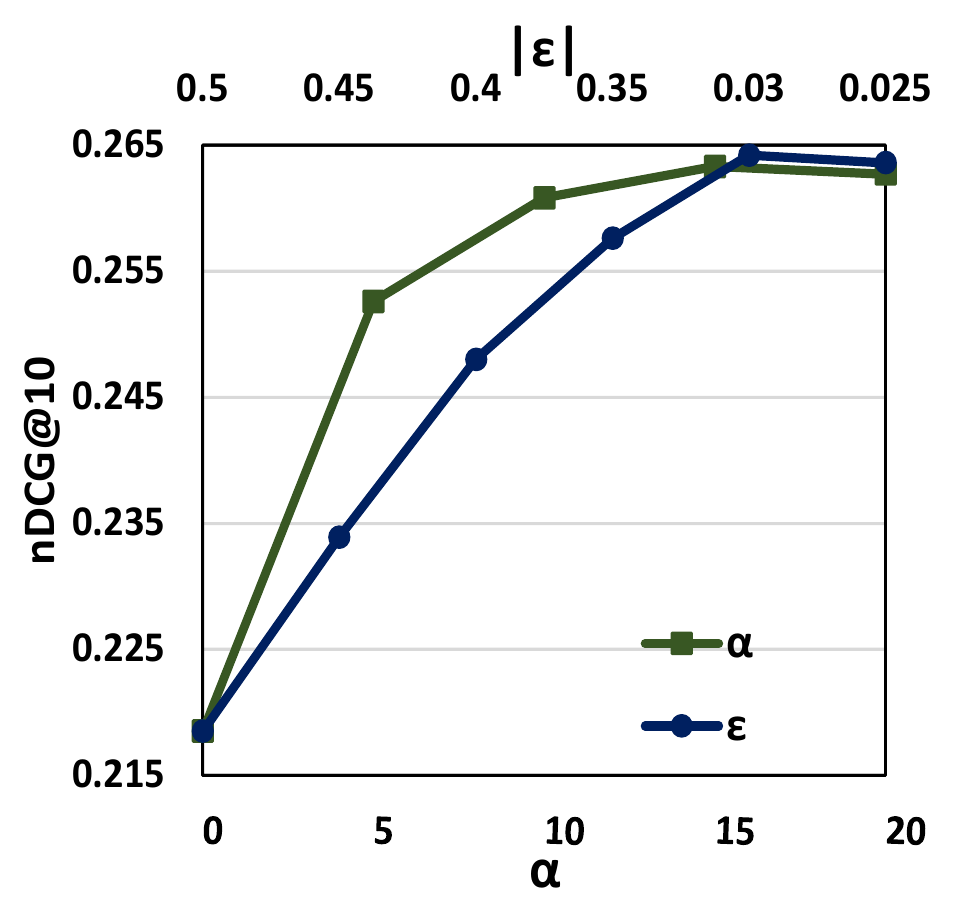} 
}
\subfigure[$x=20\%$ on CiteULike.] {  
\includegraphics[width=0.46\columnwidth]{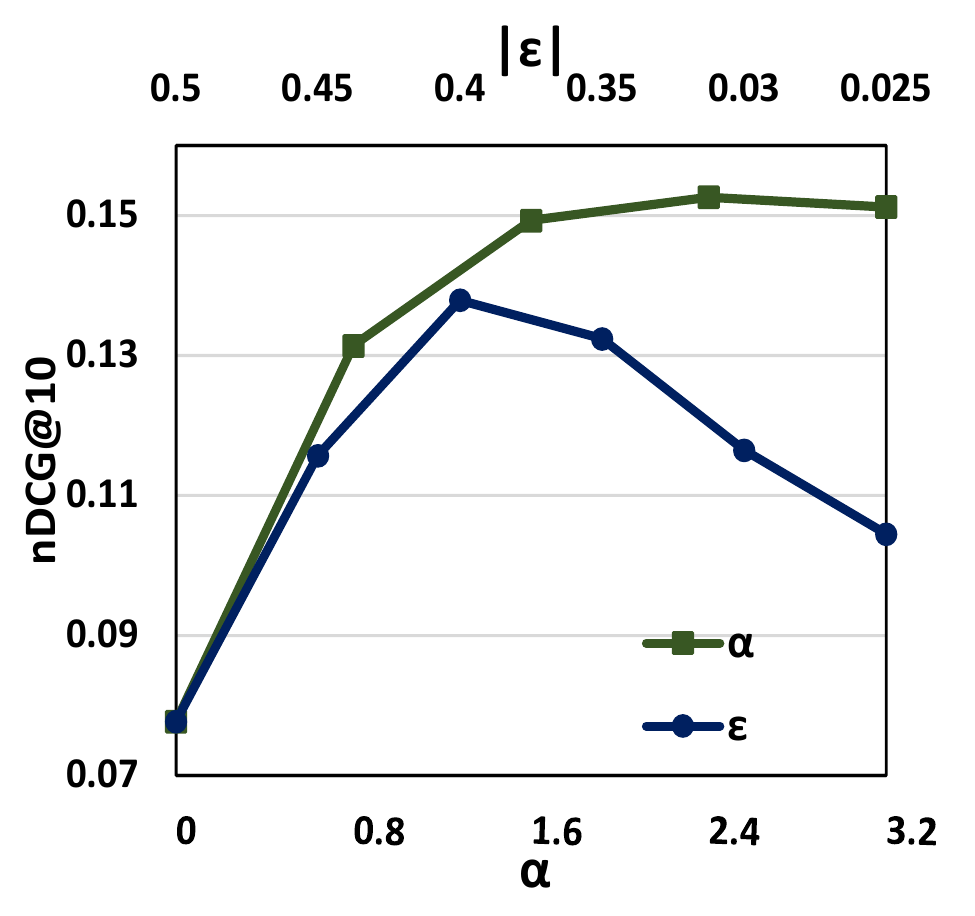} 
}
\caption{How performance changes with varying $\alpha$ and $\epsilon$.}           
\label{ggn_effectiveness}
\end{figure} 

\begin{table}[]
\caption{Comparison on training time.}
\scalebox{0.87}{
\begin{tabular}{c|cc|cc}
\toprule
\textbf{Dataset} & \multicolumn{2}{c|}{\textbf{Yelp}}                                    & \multicolumn{2}{c}{\textbf{Gowalla}}                                  \\
\textbf{Method}  & \textbf{x=80\%}                 & \textbf{x=20\%}                 & \textbf{x=80\%}                 & \textbf{x=20\%}                 \\ \midrule
DirectAU         & $3.33 \times10^3$s & $7.92\times 10^2$s & $4.53\times10^3$s & $3.30\times 10^4$s \\
LightGCN         & $1.13\times10^4$s & $2.73\times 10^3$s & $5.09\times 10^3$s & $1.22\times 10^3$s \\
GFCF             & 71.0s                             & $1.47\times10^2$s & 48.1s                             & $2.40\times 10^2$s \\
SGFCF            & 6.3s                              & 2.4s                              & 12.3s                              & 5.5s                             \\ \bottomrule
\end{tabular}}
\label{effi_compare}
\end{table}

\subsubsection{Efficiency}
We report the training time of SGFCF and several baselines that have been shown efficient in Table \ref{effi_compare}, where the results are obtained on a server equipped with AMD Ryzen 9 5950X and GeForce RTX 3090. Despite the light architecture compared with other GCN-based methods, LightGCN still requires much more training time than DirectAU whose complexity is comparable to MF. However, it requires hundreds of training epochs before convergence for training-based methods due to the non-convex loss functions, causing them to be inefficient compared with GFCF which does not require model training. Our proposed SGFCF achieves over 10x and 1000x speedup over GFCF and LightGCN, respectively. Particularly, the higher efficiency of SGFCF over GFCF is attributed to ${\rm G^2N}$. By generating a desirable spectrum via ${\rm G^2N}$, graph information is concentrated in fewer low frequency components, reducing the number of required spectral features $K$. For instance, $K=100$, 50, and 90 on CiteULike, Yelp, and Gowalla with $x=20\%$ when the accuracy is maximized, as opposed to $K=512$ on GFCF. 

\begin{figure} \centering 
\subfigure[$x=80\%$ on CiteULike.] {  
\includegraphics[width=0.46\columnwidth]{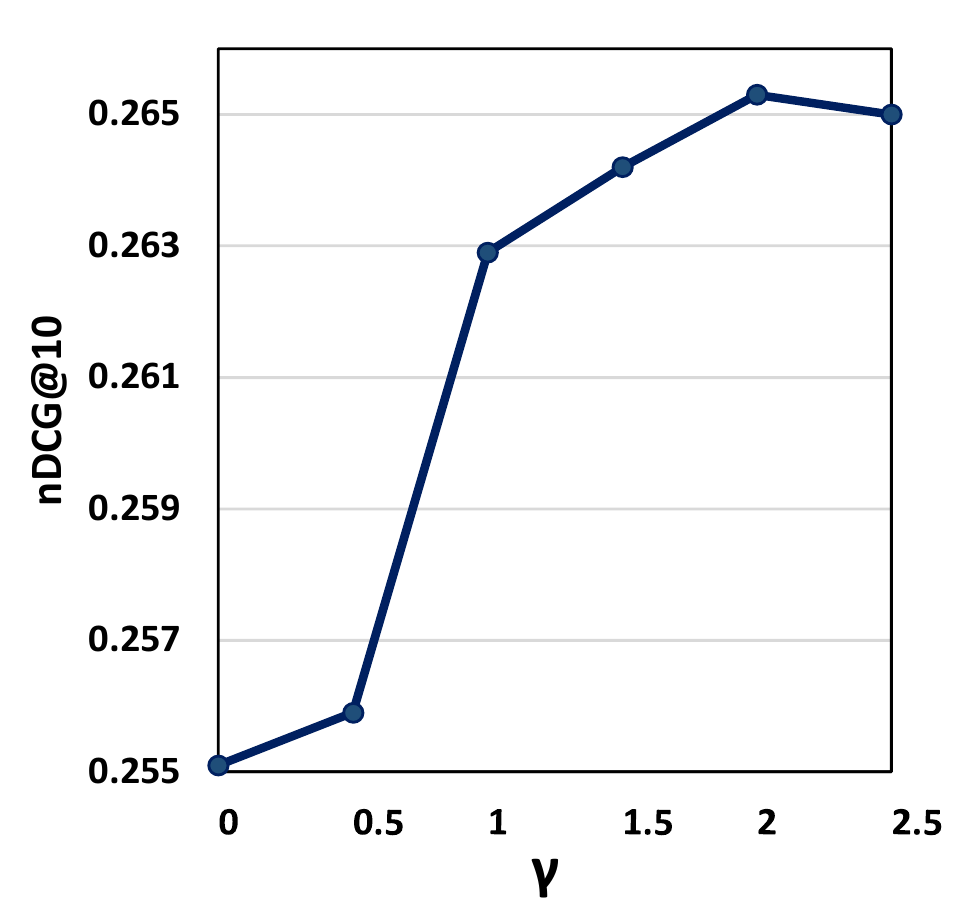} 
}
\subfigure[$x=20\%$ on CiteULike.] {  
\includegraphics[width=0.46\columnwidth]{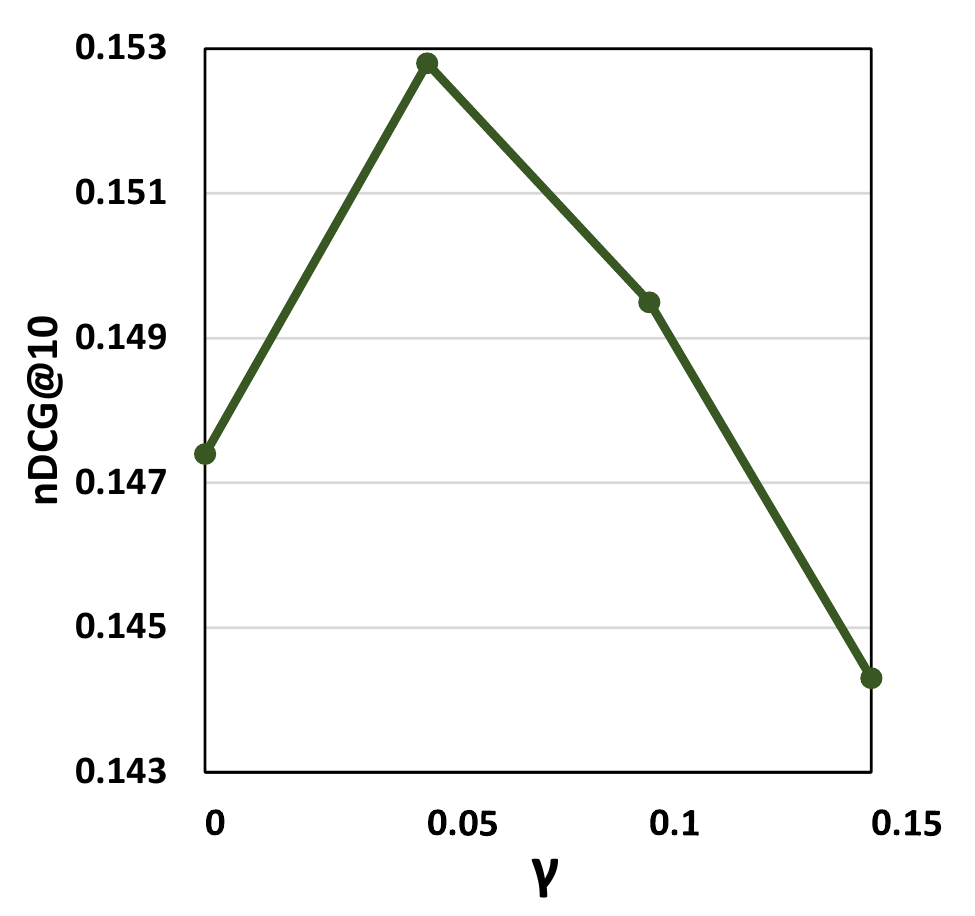} 
}

\caption{How performance changes with varying $\gamma$. }           
\label{gamma_effect_figure}
\end{figure}

\begin{figure} \centering 
\subfigure[$x=20\%$ on Gowalla.] {  
\includegraphics[width=0.46\columnwidth]{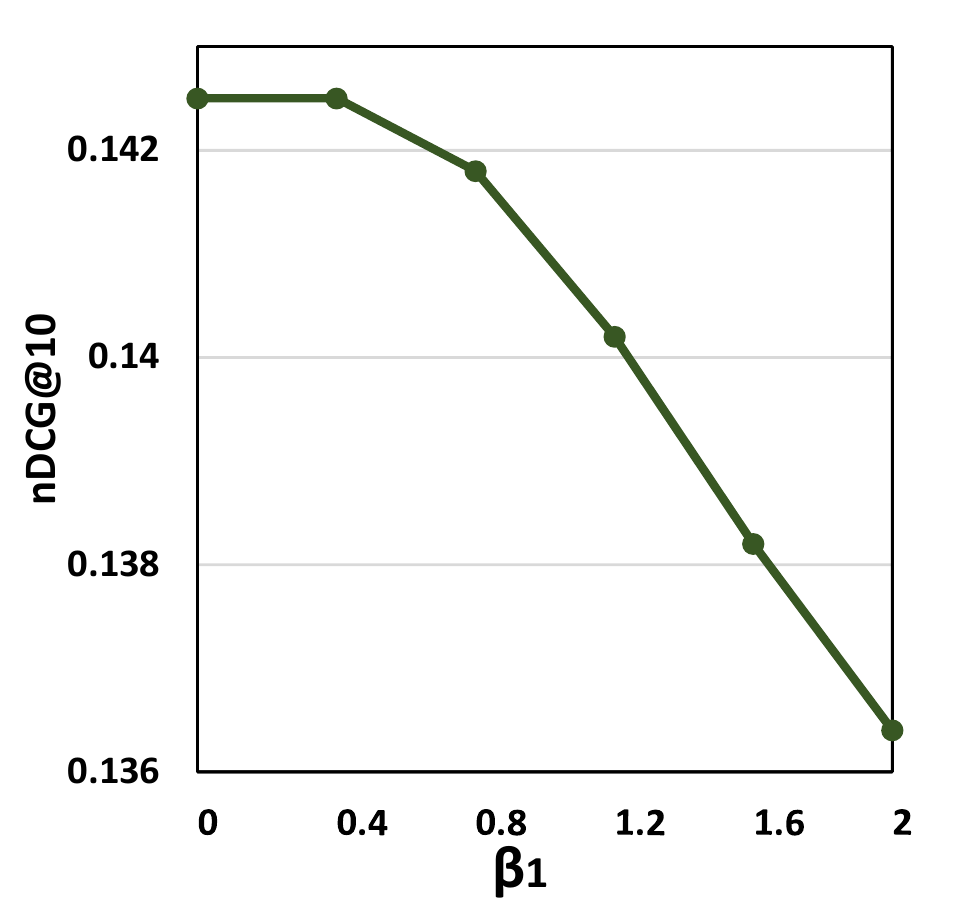} 
}
\subfigure[$x=80\%$ on Gowalla.] {  
\includegraphics[width=0.46\columnwidth]{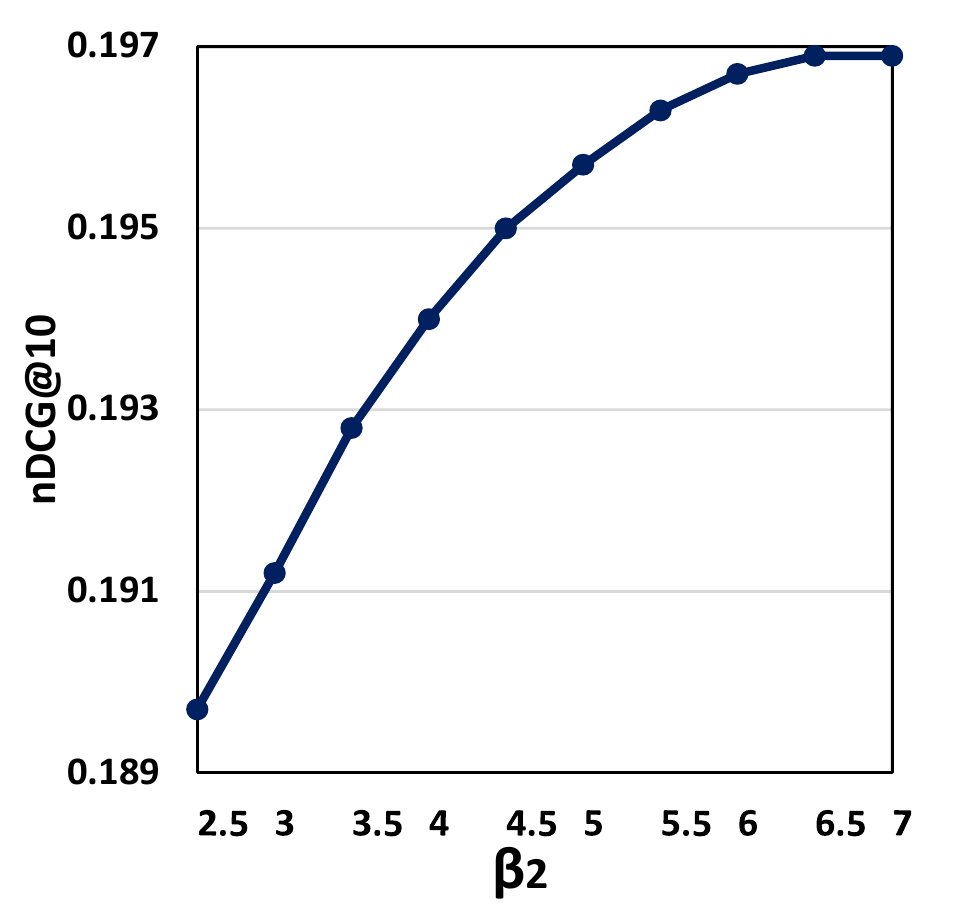} 
}

\caption{How performance changes with $\beta_1$ and $\beta_2$. }           
\label{beta_effect_figure}
\end{figure}

\subsection{Study of SGFCF}

\subsubsection{Effect of $\alpha$ and $\epsilon$}
We study how accuracy changes with $\alpha$ and $\epsilon$ and report the results in Figure \ref{ggn_effectiveness}. Since we showed that either $\alpha$ or $\epsilon$ can adjust the sharpness of spectrum in Section 4.2, we separately tune them and choose the better one, where we fix $\alpha=0$ and $\epsilon=-0.5$ when studying the other one. We can observe that the accuracy is more sensitive to $\epsilon$ than $\alpha$. Note that $\alpha \rightarrow +\infty$ and $\epsilon=0$ when nodes are equally weighted, thus $\frac{1}{\sqrt{d_u+\alpha}}$ is a smoother function than $d_u^{\epsilon}$ since $\alpha \in [0, +\infty]$ while $\epsilon\in[-0.5,0]$, explaining why $\alpha$ overall performs better than $\epsilon$. Moreover, the hyperparameter value is larger on the dense setting $x=80\%$ than the sparse setting $x=20\%$. A reasonable explanation is that nodes have smaller degrees on the sparse setting on average, on which the model performance is more sensitive to changes in hyperparameters. We can also see the difference between the values on $x=20\%$ and $x=80\%$ when the best performance is achieved is roughly consistent with their node degree difference (\textit{i.e.,} around 4 times).

\subsubsection{Effect of $\gamma$}
We study the effect of $\gamma$ and report the results in Table \ref{gamma_effect_table} and Figure \ref{gamma_effect_figure}. We can see that introducing $\gamma$ leads to a better performance, demonstrating that middle frequency components still contain some information contributing to the data representations. Particularly, the improvement on the dense setting $x=80\%$ tends to be more significant than that on the sparse setting $x=20\%$. Intuitively, the sparser data are composed of fewer spectral features. For instance, only $K=20$ spectral features are required on Yelp with $x=20\%$ when the best performance is achieved, as opposed to $K=300$ with $x=80\%$, and we observe a similar trend on other datasets. This observation implies that the middle frequencies are more noisy and useless on sparse datasets. The results in Figure \ref{gamma_effect_figure} further verifies our observation, that the accuracy is more sensitive to the change in $\gamma$ on sparse setting.  

\begin{table}[]
\caption{The improvement\% of $\gamma$ in terms of nDCG@10.}
\scalebox{1.0}{
\begin{tabular}{cclllcclllclll}
\toprule
\textbf{}     & \multicolumn{4}{c}{\textbf{Yelp}} & \textbf{CiteULike} & \multicolumn{4}{c}{\textbf{Pinterest}} & \multicolumn{4}{c}{\textbf{Gowalla}} \\
\textbf{x=80\%} & \multicolumn{4}{c}{2.35}          & 4.00               & \multicolumn{4}{c}{0.55}               & \multicolumn{4}{c}{3.66}             \\
\textbf{x=20\%} & \multicolumn{4}{c}{0.21}          & 3.66               & \multicolumn{4}{c}{0.42}               & \multicolumn{4}{c}{0.42}             \\ \bottomrule
\end{tabular}}
\label{gamma_effect_table}
\end{table}

\subsubsection{Effect of IGF} 
We show how accuracy changes with $\beta_1$ and $\beta2$ in Figure \ref{beta_effect_figure}, where we fix $\beta_2=2.0$ in (a) and $\beta_1=2.5$ in (b). We can observe that an individualized filter adapting to different users/items shows better performance than a shared graph filter.

\subsubsection{Impact of $\delta$}
Intuitively, the homophilic ratio of distinctive users/items tends to shift towards 1 as increasing $\delta$, making their difference more insignificant. Meanwhile, a larger $\delta$ also results in higher computational complexity. As shown in Table \ref{delta_effect_table}, the best performance is achieved at $\delta=2$ on most settings except the slight improvement on CiteULike with $x=20\%$. Thus, we set $\delta=2$ considering the trade-off between effectiveness and efficiency.

\section{Related Work}
\subsection{Collaborative Filtering}
Collaborative filtering is a fundamental task for recommender systems as it provides recommendations by learning from the user-item historical interactions without rely on specific knowledge or user and item profiles. The underling assumption of CF is that similar users tend to have similar preference \cite{breese1998empirical,sarwar2001item}. Matrix factorization~\cite{koren2009matrix}, one of the simplest yet effective methods for CF, characterizes users and items as learnable low-dimensional vectors, where the rating between a user and an item is estimated as the inner product between user and item vectors. Most CF methods can be considered as enhanced MF variants addressing drawbacks of MF which can be mainly classified into three categories: (1) Due to the limited available data, some works incorporate side information to help infer user preference. Rendle et al. \cite{rendle2010factorizing} introduces temporal information and combine MF with Markov chain (MC) to predict users' next behaviour. Ma et al.  \cite{ma2008sorec} integrates social relations and user-item interactions with MF. \cite{lian2014geomf} is an enhanced MF to incorporate geological information. (2) To address the drawback that MF uses a simple linear function to model complex user-item relations, much effort has been devoted to exploit advanced algorithms to learn form user-item interactions, such as multilayer perceptron~\cite{he2017neural,xue2017deep}, autoencoder~\cite{sedhain2015autorec}, attention mechanism~\cite{chen2017attentive}, transformer~\cite{sun2019bert4rec}, etc. (3) Due to the data sparsity, negative sampling is critical to generate desirable data representations. Therefore, a lot of effective sampling strategies have been proposed \cite{mao2021simplex,wang2022towards,park2023toward}.

\begin{table}[]
\caption{The effect of $\delta$ evaluated by nDCG@10.}
\scalebox{0.95}{
\begin{tabular}{ccclllcclll}
\toprule
\textbf{Settings}                & \textbf{Datasets}  & \multicolumn{4}{c}{\textbf{$\delta=2$}} & \textbf{$\delta=4$} & \multicolumn{4}{c}{\textbf{$\delta=6$}} \\ \midrule
\multirow{4}{*}{\textbf{x=80\%}} & \textbf{CiteULike} & \multicolumn{4}{c}{0.2658}              & 0.2631              & \multicolumn{4}{c}{0.2629}             \\
                                 & \textbf{Yelp}      & \multicolumn{4}{c}{0.0825}              & 0.0823              & \multicolumn{4}{c}{0.0824}             \\
                                 & \textbf{Pinterest} & \multicolumn{4}{c}{0.0919}              & 0.0919              & \multicolumn{4}{c}{0.0919}             \\
                                 & \textbf{Gowalla}   & \multicolumn{4}{c}{0.1432}              & 0.1386              & \multicolumn{4}{c}{0.1386}             \\ \midrule
\multirow{4}{*}{\textbf{x=20\%}} & \textbf{CiteULike} & \multicolumn{4}{c}{0.1542}              & 0.1547              & \multicolumn{4}{c}{0.1548}             \\
                                 & \textbf{Yelp}      & \multicolumn{4}{c}{0.0963}              & 0.0960              & \multicolumn{4}{c}{0.0953}             \\
                                 & \textbf{Pinterest} & \multicolumn{4}{c}{0.1436}              & 0.1438              & \multicolumn{4}{c}{0.1439}             \\
                                 & \textbf{Gowalla}   & \multicolumn{4}{c}{0.1971}              & 0.1951              & \multicolumn{4}{c}{0.1932}             \\ \bottomrule
\end{tabular}}
\label{delta_effect_table}
\end{table}

\subsection{GCN-based CF methods}

Graph convolution networks (GCNs) have shown great potential in recommender systems and collaborative filtering (CF). Early attempts simply apply classic GCN architectures for CF which do not necessarily show desirable performance in recommendation. For instance, SpectralCF \cite{zheng2018spectral} shares similarities with ChebNet~\cite{defferrard2016convolutional}, the designs of NGCF \cite{wang2019neural} are based on GCN \cite{kipf2017semi}, and PinSage~\cite{ying2018graph} is closely related to GraphSAGE \cite{hamilton2017inductive}. Subsequent works show the redundancy of GCN-based methods such as non-linearity and linear transformation \cite{chen2020revisiting,he2020lightgcn}, demystify how GCNs contribute to recommendation \cite{peng2022svd,peng2024less} and analyze the expressive power of GCN for recommendation \cite{shen2021powerful,cai2023expressive}. Furthermore, research effort has also be devoted to empower GCN with other advanced algorithms, such as transformer \cite{li2023graph}, sampling strategy \cite{huang2021mixgcf}, contrastive learning~\cite{wu2020self,jiang2023adaptive}, etc. and achieve further improvement.\par 

Spectral-bsed GCNs, focusing the spectral domain of graphs, have also received much attention \cite{balcilar2021analyzing,wang2022powerful}. By analyzing GCN from a perspective of graph signal processing, recent works show that GCN is essentially a low pass filter, and low/high frequencies are significantly contributive to recommendation accuracy \cite{shen2021powerful,peng2022less}. Based on this finding, several spectral GCN-based methods have been proposed and show superiority, that can be classified  into two categories: (1) non-parametric graph filters \cite{shen2021powerful,liu2023personalized} and (2) graph filters combined with supervised training \cite{peng2022less,guo2023manipulating}. However, we empirically demonstrated that they fail to perform well on datasets with different densities due to the varied noise distribution. 

\section{Conclusion}

In this work, we addressed two limitations of existing GCN-based methods: the lack of generality and expressive power. We proposed a generalized graph normalization (${\rm G^2N}$) to adjust the sharpness of spectrum, making graph filtering generalizable on datasets with different densities, and an individualized graph filtering (IGF), where we emphasize different frequencies based on the homophilic ratio measuring the distinct confidence levels of user preference that interactions can reflect, which is proved to generate arbitrary embeddings. Finally, we proposed a simplified graph filtering for CF (SGFCF) requiring only the top-$K$ singular values. Extensive experimental results on four datasets demonstrated the effectiveness and efficiency of our proposed designs. In future work, we plan to analyze the potential of GCNs from other perspectives and apply our proposed method to other recommendation tasks.\\\\

\noindent
\textbf{Acknowledgement}\\
This paper is based on results obtained from the project, “Research and Development Project of the Enhanced infrastructures for Post-5G Information and Communication Systems” (JPNP20017), commissioned by the New Energy and Industrial Technology Development Organization (NEDO).

\bibliographystyle{ACM-Reference-Format}
\balance
\bibliography{myrefs}
\newpage

\appendix

\section{Supplementary Experiments}

\begin{table}[]
\caption{Ablation study on SGFCF evaluated by nDCG@10.}
\scalebox{0.9}{
\begin{tabular}{c|clcl|clcl|clcl}
\toprule
\textbf{Dataset}                                                                       & \multicolumn{4}{c|}{\textbf{Yelp}}                                         & \multicolumn{4}{c|}{\textbf{CiteULike}}                                    & \multicolumn{4}{c}{\textbf{Gowalla}}                                      \\
\textbf{Setting}                                                                       & \multicolumn{2}{c}{\textbf{x=80\%}} & \multicolumn{2}{c|}{\textbf{x=20\%}} & \multicolumn{2}{c}{\textbf{x=80\%}} & \multicolumn{2}{c|}{\textbf{x=20\%}} & \multicolumn{2}{c}{\textbf{x=80\%}} & \multicolumn{2}{c}{\textbf{x=20\%}} \\ \midrule
SGFCF                                                                                  & \multicolumn{2}{c}{\textbf{0.0824}} & \multicolumn{2}{c|}{\textbf{0.0963}} & \multicolumn{2}{c}{\textbf{0.2667}} & \multicolumn{2}{c|}{\textbf{0.1542}} & \multicolumn{2}{c}{\textbf{0.1432}} & \multicolumn{2}{c}{\textbf{0.1973}} \\
w/o IGF                                                                                & \multicolumn{2}{c}{0.0820}          & \multicolumn{2}{c|}{0.0949}          & \multicolumn{2}{c}{0.2651}          & \multicolumn{2}{c|}{0.1532}          & \multicolumn{2}{c}{0.1388}          & \multicolumn{2}{c}{0.1922}          \\
\begin{tabular}[c]{@{}c@{}}w/o IGF \\ \& ${\rm G^2N}$\end{tabular} & \multicolumn{2}{c}{0.0537}          & \multicolumn{2}{c|}{0.0729}          & \multicolumn{2}{c}{0.2287}          & \multicolumn{2}{c|}{0.0924}          & \multicolumn{2}{c}{0.1255}          & \multicolumn{2}{c}{0.1313}          \\ \bottomrule
\end{tabular}}
\label{ablation}
\end{table} 

\begin{table}[]
\caption{Performance comparison on x=40\% and 60\%.}
\scalebox{0.73}{
\begin{tabular}{ccc|ccccc}
\toprule
\multicolumn{3}{c|}{\textbf{Setting}}                                                    & \multicolumn{5}{c}{\textbf{Methods}}                                     \\
\textbf{Dataset}                    & \multicolumn{2}{c|}{\textbf{Metric}}               & LightGCN     & GFCF         & JGCF         & SGFCF           & Improv.\% \\ \midrule
\multirow{4}{*}{\textbf{CiteULike}} & \multirow{2}{*}{\textbf{x=40\%}} & \textbf{nDCG}   & \underline{ 0.2120} & 0.2021       & 0.2095       & \textbf{0.2440} & +15.09    \\
                                    &                                  & \textbf{Recall} & \underline{ 0.1989} & 0.1951       & 0.1994       & \textbf{0.2298} & +15.54    \\
                                    & \multirow{2}{*}{\textbf{x=60\%}} & \textbf{nDCG}   & 0.2023       & \underline{ 0.2502} & 0.2039       & \textbf{0.2739} & +9.47     \\
                                    &                                  & \textbf{Recall} & 0.1969       & \underline{0.2425} & 0.1986       & \textbf{0.2622} & +8.12     \\ \midrule
\multirow{4}{*}{\textbf{Pinterest}} & \multirow{2}{*}{\textbf{x=40\%}} & \textbf{nDCG}   & 0.1204       & 0.1275       & \underline{ 0.1294} & \textbf{0.1371} & +5.95     \\
                                    &                                  & \textbf{Recall} & 0.1170       & 0.1221       & \underline{ 0.1241} & \textbf{0.1313} & +5.80     \\
                                    & \multirow{2}{*}{\textbf{x=60\%}} & \textbf{nDCG}   & 0.0912       & \underline{ 0.1064} & 0.1025       & \textbf{0.1117} & +4.98     \\
                                    &                                  & \textbf{Recall} & 0.0891       & \underline{ 0.1022} & 0.0990       & \textbf{0.1067} & +4.40     \\ \bottomrule
\end{tabular}}
\label{compare_2}
\end{table}

\begin{table}[]
\caption{Performance (nDCG@10) on different graph filters.}
\scalebox{0.88}{
\begin{tabular}{c|clcl|clcl|clcl}
\toprule
\textbf{Dataset} & \multicolumn{4}{c|}{\textbf{Yelp}}                                         & \multicolumn{4}{c|}{\textbf{Pinterest}}                                    & \multicolumn{4}{c}{\textbf{Gowalla}}                                      \\
\textbf{Setting} & \multicolumn{2}{c}{\textbf{x=80\%}} & \multicolumn{2}{c|}{\textbf{x=20\%}} & \multicolumn{2}{c}{\textbf{x=80\%}} & \multicolumn{2}{c|}{\textbf{x=20\%}} & \multicolumn{2}{c}{\textbf{x=80\%}} & \multicolumn{2}{c}{\textbf{x=20\%}} \\ \midrule
Monomial         & \multicolumn{2}{c}{\textbf{0.0820}} & \multicolumn{2}{c|}{\textbf{0.0949}}          & \multicolumn{2}{c}{\textbf{0.0922}} & \multicolumn{2}{c|}{\textbf{0.1435}} & \multicolumn{2}{c}{\textbf{0.1388}} & \multicolumn{2}{c}{\textbf{0.1922}} \\
Exp              & \multicolumn{2}{c}{0.0816}          & \multicolumn{2}{c|}{0.0940}          & \multicolumn{2}{c}{0.0919} & \multicolumn{2}{c|}{0.1432}          & \multicolumn{2}{c}{0.1385}          & \multicolumn{2}{c}{\textbf{0.1922}} \\
Markov           & \multicolumn{2}{c}{0.0802}          & \multicolumn{2}{c|}{0.0945} & \multicolumn{2}{c}{0.0904}          & \multicolumn{2}{c|}{0.1406}          & \multicolumn{2}{c}{0.1335}          & \multicolumn{2}{c}{0.1714}          \\
Jacobi            & \multicolumn{2}{c}{0.0770}          & \multicolumn{2}{c|}{0.0784}          & \multicolumn{2}{c}{0.0856}          & \multicolumn{2}{c|}{0.1425}          & \multicolumn{2}{c}{0.1381}          & \multicolumn{2}{c}{0.1909}          \\ \bottomrule
\end{tabular}}
\label{filter_choice}
\end{table}

\subsection{Ablation Study}
To demonstrate the effectiveness of our proposed designs, we compare three variants: (1) SGFCF, (2) SGFCF without IGF, and (3) SGFCF without both ${\rm G^2N}$ and IGF, and report the results in Table \ref{ablation}. We observe that removing either of ${\rm G^2N}$ and IGF results in performance degradation, showing that both ${\rm G^2N}$ and IGF are contributive to model performance. Particularly, IGF tends to be more effective on the sparse setting, indicating that it might require stronger expressive power to generate the optimal representations on the sparse data. Moreover, ${\rm G^2N}$ contributes more to the accuracy than IGF, as the poor performance of graph filters is mainly due to the varied noise distribution polluting the low frequencies being important to the data representations.

\subsection{Experiments on Other Density Settings}
To further verify the effectiveness of our methods, we compare our SGFCF with several competitive baselines on x=40\% and 60\%, and report the results in Table \ref{compare_2}. We can observe that GFCF without training shows better performance on the denser setting x=60\%, while training-based methods JGCF and LightGCN achieve better results on the sparser setting x=40\% as supervised training shows superior ability learning from noisy data. Our proposed SGFCF outperforms all baselines across the board, demonstrating the generality of our proposed designs.

\subsection{Graph Filter Designs}
We compare four commonly used graph filters for recommendation: monomial filter, exponential diffusion kernel, Markov diffusion kernel, and Jacobi polynomials (with detailed introduction as follows) and report the results in Table~\ref{filter_choice}. Overall, the monomial filter: $\lambda_k^{\beta}$ with the simplest design shows better performance than other filters. Due to the important features that are concentrated in only a few low frequencies via ${\rm G^2N}$, a simple increasing function is already able to appropriately emphasize different frequencies according to their importance instead of complicated band-pass filters such as the Markov diffusion kernel and Jacobi polynomials.
\subsubsection{Monomial Filter}
The eigenvalue of $\mathbf{\hat{A}}^l$ is $\lambda_k^l$, here we extend it to $\lambda_k^{\beta}$ where $\beta\geq0$. It is a simple increasing function.

\subsubsection{Exponential Diffusion Kernel}
The exponential diffusion kernel is defined as:
\begin{equation}
\exp(\mathbf{\beta \hat{A}})=\sum_{l=0}^{\infty}\frac{\beta^l \mathbf{\hat{A}}^l}{l!},
\end{equation}
where the corresponding eigenvalue is $e^{\beta \lambda_k}$.

\subsubsection{Markov Diffusion Kernel}
The Markov diffusion kernel is defined as: $\frac{1}{L} \sum_{l=0}^L \mathbf{\hat{A}}^l$ where the corresponding eigenvalue is $\frac{\sum_{l=0}^L\lambda_k^l}{L}$.

\subsubsection{Jacobi Polynomials}
The Jacobi basis for $k\geq2$ is defined as:

\begin{equation}
P^{a,b}_k(\mathbf{\hat{A}})\mathbf{E}=\theta_k\mathbf{\hat{A}}P^{a,b}_{k-1}(\mathbf{\hat{A}})\mathbf{E}+\theta_k'P^{a,b}_{k-1}(\mathbf{\hat{A}})\mathbf{E}-\theta_k''P^{a,b}_{k-2}(\mathbf{\hat{A}})\mathbf{E},
\end{equation}
where
\begin{equation}
\begin{aligned}
&\theta_k=\frac{(2k+a+b)(2k+a+b-1)}{2k(k+a+b)},\\
&\theta_k'=\frac{(2k+a+b-1)(a^2-b^2)}{2k(k+a+b)(2k+a+b-2)},\\
& \theta_k''=\frac{(k+a-1)(k+b-1)(2k+a+b)}{k(k+a+b)(2k+a+b-2)}.
\end{aligned}
\end{equation}
For $k<2$,
\begin{equation}
\begin{aligned}
&P^{a,b}_0(\mathbf{\hat{A}})\mathbf{E}=\mathbf{E},\\
&P^{a,b}_0(\mathbf{\hat{A}})\mathbf{E}=\mathbf{E}=\frac{a-b}{2}\mathbf{E}+\frac{a+b+2}{2}\mathbf{\hat{A}}\mathbf{E}.
\end{aligned}
\end{equation}
The final representations are generated by summing up the embeddings from each iteration: $\sum_{k=0}^L P^{a,b}_k(\mathbf{\hat{A}})$.

\section{Proofs}

\subsection{Proof of Theorem 1 and Corollary 1}
\begin{proof}
Let $\mathbf{p}_k$, $\mathbf{q}_k$, and $\sigma_k$ be the left, right singular vector, and the singular value of $\mathbf{\hat{R}}$, we can show the following relations:
\begin{equation}
\begin{aligned}
&\begin{bmatrix}
 \mathbf{0}& \mathbf{\hat{R}}\\ 
\mathbf{\hat{R}}^T &\mathbf{0} 
\end{bmatrix}
\begin{bmatrix}
\mathbf{p}_k\\
\mathbf{q}_k\\
\end{bmatrix}
=
\begin{bmatrix}
\mathbf{\hat{R}}\mathbf{q}_k\\
\mathbf{\hat{R}}^T\mathbf{p}_k\\
\end{bmatrix}
=\sigma_k
\begin{bmatrix}
\mathbf{p}_k\\
\mathbf{q}_k\\
\end{bmatrix},\\
&\begin{bmatrix}
 \mathbf{0}& \mathbf{\hat{R}}\\ 
\mathbf{\hat{R}}^T &\mathbf{0} 
\end{bmatrix}
\begin{bmatrix}
\mathbf{p}_k\\
-\mathbf{q}_k\\
\end{bmatrix}
=
\begin{bmatrix}
-\mathbf{\hat{R}}\mathbf{q}_k\\
\mathbf{\hat{R}}^T\mathbf{p}_k\\
\end{bmatrix}
=-\sigma_k
\begin{bmatrix}
\mathbf{p}_k\\
\mathbf{q}_k\\
\end{bmatrix},
\end{aligned}
\end{equation}
where $\sigma_k$ and $-\sigma_k$ are two eigenvalues of $\mathbf{\hat{A}}$, with $[\mathbf{p}_k,\mathbf{q}_k]$ and $[\mathbf{p}_k,-\mathbf{q}_k]$ as the corresponding eigenvectors, respectively. According to Theorem 1, it is easy to show $r_{ui}^{(K)}= r_{ui}^{(n-K)}=\mathbf{P}^{(K)}{\mathbf{Q}^{(K)}}^T$. Corollary 1 implies that the rating estimated by the high frequencies are exactly opposite to the symmetric (to $\lambda_{n/2}=0$) low frequencies based on Equation (\ref{uniform_filter}).
\end{proof}

\subsection{Proof of Theorem 2}
\begin{proof}
We first consider $d=1$. If LGCN can generate arbitrary embeddings, then the following Equation holds:
\begin{equation}
\mathbf{V}^T\mathbf{O}=diag\left(g(\lambda_k) \right) \mathbf{V}^T \mathbf{E},
\end{equation} 
which is equivalent to the following Equation:
\begin{equation}
\begin{bmatrix}
\frac{\left(\mathbf{V}^T\mathbf{O}\right)_1}{\left(\mathbf{V}^T\mathbf{E}\right)_1}\\
\vdots\\
\frac{\left(\mathbf{V}^T\mathbf{O}\right)_n}{\left(\mathbf{V}^T\mathbf{E}\right)_n}
\end{bmatrix}
=
\begin{bmatrix}
\sum_l \theta_l \lambda^l_1\\
\vdots\\
\sum_l \theta_l \lambda^l_n
\end{bmatrix}
=\mathbf{B}
\begin{bmatrix}
\theta_1\\
\vdots\\
\theta_n
\end{bmatrix},
\label{thoerm2_proof}
\end{equation}
where $\mathbf{B}\in \mathbb{R}^{n\times L}$, $\mathbf{B}_{kl}=\lambda_k^l$. By extending the order of $g(\lambda_k)$ to $n$ (\textit{i.e.,} $L=n$), $\mathbf{B}$ becomes a full rank matrix as long as  $\mathbf{\hat{A}}$ has no repeated eigenvalue, thus we can always find a solution to satisfy Equation (\ref{thoerm2_proof}). To generalize the above result to the situation of $d>1$, multiple filters are required:
\begin{equation}
\begin{bmatrix}
\frac{\left(\mathbf{V}^T\mathbf{O}\right)_{11}}{\left(\mathbf{V}^T\mathbf{E}\right)_{11}},\cdots,
\frac{\left(\mathbf{V}^T\mathbf{O}\right)_{1d}}{\left(\mathbf{V}^T\mathbf{E}\right)_{1d}} \\
\vdots \qquad \qquad \quad \vdots\\
\frac{\left(\mathbf{V}^T\mathbf{O}\right)_{n1}}{\left(\mathbf{V}^T\mathbf{E}\right)_{n1}},\cdots,
\frac{\left(\mathbf{V}^T\mathbf{O}\right)_{nd}}{\left(\mathbf{V}^T\mathbf{E}\right)_{nd}}
\end{bmatrix}
=\mathbf{B}
\begin{bmatrix}
\theta_{11},\cdots,\theta_{1d}\\
\vdots\\
\theta_{n1},\cdots,\theta_{nd}
\end{bmatrix},
\label{thoerm2_proof1}
\end{equation}
By applying Equation (\ref{thoerm2_proof1}) to LGCN, we get Equation (\ref{modified_lgcn}). 
\end{proof}

\subsection{Proof of Theorem 3}
\begin{proof}
We first show that the Frobenius norm can be defined as follows:
\begin{equation}
\left|\mathbf{\hat{A}}\right|_F^2=\mathbf{trace}\left( \mathbf{\hat{A}}^T \mathbf{\hat{A}}  \right)=\sum_{k=1}^n \lambda_k^2
\end{equation}
According to Theorem 1, $\sum_{k=1}^{n/2} \lambda_k^2=\sum_{k=n/2}^n \lambda_k^2$ where $\lambda_{n/2}=0$. Let $\frac{\tilde{\lambda}^2_k}{\lambda^2_k}=\alpha_k$, then $\alpha_1\geq\cdots\geq\alpha_{K}>\alpha_{K+1}\geq\cdots \geq \alpha_{n/2}$, and the following relation holds:
\begin{equation}
\begin{aligned}
&2\frac{\left|\mathbf{\tilde{A}}_K \right|_F^2}{\left|\mathbf{\tilde{A}} \right|_F^2}
=\frac{\sum_{k=1}^K \tilde{\lambda}^2_k}{\sum_{k=1}^{n/2} \tilde{\lambda}^2_k}=\frac{\sum_{k=1}^K \alpha_k \lambda^2_k}{\sum_{k=1}^K \alpha_k \lambda_k^2 + \sum_{j=K+1}^{n/2} \alpha_j \lambda^2_j}\\
&\geq \frac{ \alpha_K \sum_{k=1}^K \lambda^2_k}{ \alpha_K \sum_{k=1}^K \lambda_k + {\sum_{k=1}^K \alpha_k \lambda_k^2 + \sum_{j=K+1}^{n/2} \alpha_j \lambda^2_j}}\\
&\geq \frac{ \alpha_K \sum_{k=1}^K \lambda^2_k}{ \alpha_K \sum_{k=1}^K \lambda_k + \alpha_{K+1} \sum_{j=K+1}^{n/2} \lambda^2_j}\\
&> \frac{\sum_{k=1}^K \lambda^2_k}{\sum_{k=1}^{n/2} \lambda^2_k}=2\frac{\left|\mathbf{\hat{A}}_K \right|_F^2}{\left|\mathbf{\hat{A}} \right|_F^2}
\end{aligned}
\end{equation}
The inequalities above are based on the observation that $\frac{x}{x+y}=1-\frac{y}{x+y}$ decreases as $x$ decreases and $y$ increases for $x>0$ and $y>0$.
\end{proof}

\subsection{Proof of Theorem 4}
\begin{proof}
According to Courant-Fischer Theorem \cite{spielman2012spectral}, we have:
\begin{equation}
\lambda_k= \mathop{max}_{dim(S)=k} \mathop{min}_{x\in S} \mathbf{x}^T \mathbf{\hat{A}}\mathbf{x}\qquad s.t. \left|\mathbf{x}\right|=1
\end{equation}
where the eigenvalue is arranged in descending order. Then,
\begin{equation}
\begin{aligned}
&\tilde{\lambda}_k= \mathop{max}_{dim(S)=k} \mathop{min}_{x\in S} \mathbf{x}^T \mathbf{\tilde{A}}\mathbf{x}=\mathop{max}_{dim(S)=k} \mathop{min}_{x\in S} \sum_{(u,i)\in\mathcal{E}} \frac{2\mathbf{x}_u \mathbf{x}_i}{(d_u+\alpha)^{{\mbox -}\epsilon}(d_i+\alpha)^{{\mbox -}\epsilon}}\\
&=\mathop{max} \mathop{min} \sum_{(u,i)\in\mathcal{E}} \frac{2\mathbf{x}_u \mathbf{x}_i}{(d_u)^{0.5}(d_i)^{0.5}}\sqrt{\frac{d_u d_i}{(d_u+\alpha)(d_i+\alpha)}}\left(\left(d_u+\alpha \right)\left(d_i+\alpha \right)\right)^{0.5+\epsilon}\\
&\leq d_{max}\left(d_{max}+\alpha \right)^{2\epsilon} \left( \mathop{max} \mathop{min} \mathbf{x}^T \mathbf{\hat{A}}\mathbf{x} \right)= d_{max}\left(d_{max}+\alpha \right)^{2\epsilon} \lambda_k.
\end{aligned}
\end{equation}
In the second last step, $\frac{d_u}{d_u+\alpha}$ increases over $d_u$, thus $\sqrt{\frac{d_u d_i}{(d_u+\alpha)(d_i+\alpha)}} \leq \frac{d_{max}}{d_{max}+\alpha}$. It is evident that $\left(\left(d_u+\alpha \right)\left(d_i+\alpha \right)\right)^{0.5+\epsilon}\leq \left( d_{max}+\alpha \right)^{1+2\epsilon}$. Similarly, we can prove $\tilde{\lambda}_k\geq d_{min}\left(d_{min}+\alpha \right)^{2\epsilon} \lambda_k$.

\end{proof}

\subsection{Proof of Theorem 5}
\begin{proof}
The power of the adjacency matrix can be rewritten as follows:
\begin{equation}
\mathbf{\hat{A}}^l=
\left\{
             \begin{array}{lr}
             \begin{bmatrix}
 			\left(\mathbf{\hat{R}}\mathbf{\hat{R}}^T\right)^{\frac{l}{2}}&\mathbf{0}\\ 
			\mathbf{0} & \left(\mathbf{\hat{R}}^T\mathbf{\hat{R}}\right)^{\frac{l}{2}}
			\end{bmatrix}  \quad\quad\quad\quad\,\, l=\{0, 2, 4, \cdots \}    \\\\
             
             \begin{bmatrix}
 			\mathbf{0}& \mathbf{\hat{R}}\left(\mathbf{\hat{R}}^T\mathbf{\hat{R}}
 			\right)^{\frac{l\mbox{-}1}{2}}\\
			\mathbf{\hat{R}}^T\left(\mathbf{\hat{R}}\mathbf{\hat{R}}^T\right)^{\frac{l\mbox{-}1}{2}} &
			\mathbf{0}
			\end{bmatrix}   \quad   l=\{1, 3, 5, \cdots \}.   \\
             \end{array}
\right.
\end{equation}
According to singular value decomposition (SVD): $\mathbf{\hat{R}}=\mathbf{P}diag(\lambda_k)\mathbf{Q}^T$, it is easy to show the following relations:
\begin{equation}
\begin{aligned}
&\left(\mathbf{\hat{R}}\mathbf{\hat{R}}^T\right)^{\frac{l}{2}}=\mathbf{P}diag\left(\sigma_k^{l}\right)\mathbf{P}^T,\quad
\left(\mathbf{\hat{R}}^T\mathbf{\hat{R}}\right)^{\frac{l}{2}}=\mathbf{Q}diag\left(\sigma_k^{l}\right)\mathbf{Q}^T,\\
&\mathbf{\hat{R}}\left(\mathbf{\hat{R}}^T\mathbf{\hat{R}}\right)^{\frac{l\mbox{-}1}{2}}=\mathbf{P}diag\left(\sigma_k^{l}\right)\mathbf{Q}^T,\quad
\mathbf{\hat{R}}^T\left(\mathbf{\hat{R}}\mathbf{\hat{R}}^T\right)^{\frac{l\mbox{-}1}{2}}=\mathbf{Q}diag\left(\sigma_k^{l}\right)\mathbf{P}^T.
\end{aligned}
\end{equation}
Then, the final embeddings of SGF can be formulated as follows:
\begin{equation}
\mathbf{O}=\sum_{l=0}^L\mathbf{\hat{A}}^l=\begin{bmatrix}
 \mathbf{P}diag(\psi(\sigma_k)) \mathbf{P}^T& \mathbf{P}diag(\omega(\sigma_k))\mathbf{Q}^T\\ 
\mathbf{Q}diag(\omega(\sigma_k))\mathbf{P}^T &\mathbf{Q}diag(\psi(\sigma_k))\mathbf{Q}^T
\end{bmatrix},
\end{equation}
where $\psi(\sigma_k)=\sum_{l=\{0, 2, \cdots \}}\sigma_k^l$, $\omega(\sigma_k)=\sum_{l=\{1, 3, \cdots \}}\sigma_k^l$. Then, the ratings are estimated as:
\begin{equation}
\mathbf{O}_U \mathbf{O}_I^T=\mathbf{P}diag(\phi(\lambda_k)\omega(\lambda_k))\mathbf{Q}^T
\end{equation}
\end{proof}

\end{document}